\documentclass[lettersize,10pt,journal]{IEEEtran}

\IEEEoverridecommandlockouts

\usepackage{cite}
\usepackage[T1]{fontenc}
\usepackage{graphicx}
\usepackage{amssymb}
\usepackage{amsmath}
\usepackage{amsthm}
\usepackage{microtype} 
\usepackage{balance}
\usepackage{subfigure}
\usepackage{xcolor} 
\usepackage{booktabs} 
\usepackage{dsfont}

\usepackage{amsfonts}
\usepackage{algorithm}
\usepackage{algorithmic}
\usepackage{graphicx}
\usepackage{textcomp}
\usepackage{xcolor}
\usepackage{microtype}
\usepackage{subfigure} 
\usepackage{adjustbox} 
\usepackage{balance}

\usepackage{tablefootnote}

\usepackage{url}

\usepackage{amssymb}
\usepackage{amsfonts}
\usepackage{mathrsfs}
\usepackage{xspace}
\usepackage{bm}
\usepackage{upgreek}

\newcommand{\safemath}[2]{\newcommand{#1}{\ensuremath{#2}\xspace}}

\safemath{\bma}{\mathbf{a}}
\safemath{\bmb}{\mathbf{b}}
\safemath{\bmc}{\mathbf{c}}
\safemath{\bmd}{\mathbf{d}}
\safemath{\bme}{\mathbf{e}}
\safemath{\bmf}{\mathbf{f}}
\safemath{\bmg}{\mathbf{g}}
\safemath{\bmh}{\mathbf{h}}
\safemath{\bmi}{\mathbf{i}}
\safemath{\bmj}{\mathbf{j}}
\safemath{\bmk}{\mathbf{k}}
\safemath{\bml}{\mathbf{l}}
\safemath{\bmm}{\mathbf{m}}
\safemath{\bmn}{\mathbf{n}}
\safemath{\bmo}{\mathbf{o}}
\safemath{\bmp}{\mathbf{p}}
\safemath{\bmq}{\mathbf{q}}
\safemath{\bmr}{\mathbf{r}}
\safemath{\bms}{\mathbf{s}}
\safemath{\bmt}{\mathbf{t}}
\safemath{\bmu}{\mathbf{u}}
\safemath{\bmv}{\mathbf{v}}
\safemath{\bmw}{\mathbf{w}}
\safemath{\bmx}{\mathbf{x}}
\safemath{\bmy}{\mathbf{y}}
\safemath{\bmz}{\mathbf{z}}
\safemath{\bmzero}{\mathbf{0}}
\safemath{\bmone}{\mathbf{1}}

\bmdefine{\biad}{a}
\bmdefine{\bibd}{b}
\bmdefine{\bicd}{c}
\bmdefine{\bidd}{d}
\bmdefine{\bied}{e}
\bmdefine{\bifd}{f}
\bmdefine{\bigd}{g}
\bmdefine{\bihd}{h}
\bmdefine{\biid}{i}
\bmdefine{\bijd}{j}
\bmdefine{\bikd}{k}
\bmdefine{\bild}{l}
\bmdefine{\bimd}{m}
\bmdefine{\bind}{n}
\bmdefine{\biod}{o}
\bmdefine{\bipd}{p}
\bmdefine{\biqd}{q}
\bmdefine{\bird}{r}
\bmdefine{\bisd}{s}
\bmdefine{\bitd}{t}
\bmdefine{\biud}{u}
\bmdefine{\bivd}{v}
\bmdefine{\biwd}{w}
\bmdefine{\bixd}{x}
\bmdefine{\biyd}{y}
\bmdefine{\bizd}{z}

\bmdefine{\bixid}{\xi}
\bmdefine{\bilambdad}{\lambda}
\bmdefine{\bimud}{\mu}
\bmdefine{\bithetad}{\theta}
\bmdefine{\biphid}{\phi}
\bmdefine{\bideltad}{\delta}

\safemath{\bmia}{\biad}
\safemath{\bmib}{\bibd}
\safemath{\bmic}{\bicd}
\safemath{\bmid}{\bidd}
\safemath{\bmie}{\bied}
\safemath{\bmif}{\bifd}
\safemath{\bmig}{\bigd}
\safemath{\bmih}{\bihd}
\safemath{\bmii}{\biid}
\safemath{\bmij}{\bijd}
\safemath{\bmik}{\bikd}
\safemath{\bmil}{\bild}
\safemath{\bmim}{\bimd}
\safemath{\bmin}{\bind}
\safemath{\bmio}{\biod}
\safemath{\bmip}{\bipd}
\safemath{\bmiq}{\biqd}
\safemath{\bmir}{\bird}
\safemath{\bmis}{\bisd}
\safemath{\bmit}{\bitd}
\safemath{\bmiu}{\biud}
\safemath{\bmiv}{\bivd}
\safemath{\bmiw}{\biwd}
\safemath{\bmix}{\bixd}
\safemath{\bmiy}{\biyd}
\safemath{\bmiz}{\bizd}

\safemath{\bmxi}{\bixid}
\safemath{\bmlambda}{\bilambdad}
\safemath{\bmmu}{\bimud}
\safemath{\bmtheta}{\bithetad}
\safemath{\bmphi}{\biphid}
\safemath{\bmdelta}{\bideltad}

\safemath{\bA}{\mathbf{A}}
\safemath{\bB}{\mathbf{B}}
\safemath{\bC}{\mathbf{C}}
\safemath{\bD}{\mathbf{D}}
\safemath{\bE}{\mathbf{E}}
\safemath{\bF}{\mathbf{F}}
\safemath{\bG}{\mathbf{G}}
\safemath{\bH}{\mathbf{H}}
\safemath{\bI}{\mathbf{I}}
\safemath{\bJ}{\mathbf{J}}
\safemath{\bK}{\mathbf{K}}
\safemath{\bL}{\mathbf{L}}
\safemath{\bM}{\mathbf{M}}
\safemath{\bN}{\mathbf{N}}
\safemath{\bO}{\mathbf{O}}
\safemath{\bP}{\mathbf{P}}
\safemath{\bQ}{\mathbf{Q}}
\safemath{\bR}{\mathbf{R}}
\safemath{\bS}{\mathbf{S}}
\safemath{\bT}{\mathbf{T}}
\safemath{\bU}{\mathbf{U}}
\safemath{\bV}{\mathbf{V}}
\safemath{\bW}{\mathbf{W}}
\safemath{\bX}{\mathbf{X}}
\safemath{\bY}{\mathbf{Y}}
\safemath{\bZ}{\mathbf{Z}}

\safemath{\bZero}{\mathbf{0}}
\safemath{\bOne}{\mathbf{1}}
\safemath{\bDelta}{\mathbf{\Delta}}
\safemath{\bLambda}{\mathbf{\UpLambda}}
\safemath{\bPhi}{\mathbf{\Upphi}}
\safemath{\bSigma}{\mathbf{\Upsigma}}
\safemath{\bOmega}{\mathbf{\Upomega}}
\safemath{\bTheta}{\mathbf{\Uptheta}}

\bmdefine{\biAd}{A}
\bmdefine{\biBd}{B}
\bmdefine{\biCd}{C}
\bmdefine{\biDd}{D}
\bmdefine{\biEd}{E}
\bmdefine{\biFd}{F}
\bmdefine{\biGd}{G}
\bmdefine{\biHd}{H}
\bmdefine{\biId}{I}
\bmdefine{\biJd}{J}
\bmdefine{\biKd}{K}
\bmdefine{\biLd}{L}
\bmdefine{\biMd}{M}
\bmdefine{\biOd}{N}
\bmdefine{\biPd}{O}
\bmdefine{\biQd}{P}
\bmdefine{\biRd}{R}
\bmdefine{\biSd}{S}
\bmdefine{\biTd}{T}
\bmdefine{\biUd}{U}
\bmdefine{\biVd}{V}
\bmdefine{\biWd}{W}
\bmdefine{\biXd}{X}
\bmdefine{\biYd}{Y}
\bmdefine{\biZd}{Z}

\bmdefine{\biDelta}{\Delta}
\bmdefine{\biLambda}{\Lambda}
\bmdefine{\biPhi}{\Phi}
\bmdefine{\biSigma}{\Sigma}
\bmdefine{\biOmega}{\Omega}
\bmdefine{\biTheta}{\Theta}

\safemath{\bimA}{\biAd}
\safemath{\bimB}{\biBd}
\safemath{\bimC}{\biCd}
\safemath{\bimD}{\biDd}
\safemath{\bimE}{\biEd}
\safemath{\bimF}{\biFd}
\safemath{\bimG}{\biGd}
\safemath{\bimH}{\biHd}
\safemath{\bimI}{\biId}
\safemath{\bimJ}{\biJd}
\safemath{\bimK}{\biKd}
\safemath{\bimL}{\biLd}
\safemath{\bimM}{\biMd}
\safemath{\bimN}{\biNd}
\safemath{\bimO}{\biOd}
\safemath{\bimP}{\biPd}
\safemath{\bimQ}{\biQd}
\safemath{\bimR}{\biRd}
\safemath{\bimS}{\biSd}
\safemath{\bimT}{\biTd}
\safemath{\bimU}{\biUd}
\safemath{\bimV}{\biVd}
\safemath{\bimW}{\biWd}
\safemath{\bimX}{\biXd}
\safemath{\bimY}{\biYd}
\safemath{\bimZ}{\biZd}

\safemath{\bimDelta}{\biDelta}
\safemath{\bimLambda}{\biLambda}
\safemath{\bimPhi}{\biPhi}
\safemath{\bimSigma}{\biSigma}
\safemath{\bimOmega}{\biOmega}
\safemath{\bimTheta}{\biTheta}

\safemath{\setA}{\mathcal{A}}
\safemath{\setB}{\mathcal{B}}
\safemath{\setC}{\mathcal{C}}
\safemath{\setD}{\mathcal{D}}
\safemath{\setE}{\mathcal{E}}
\safemath{\setF}{\mathcal{F}}
\safemath{\setG}{\mathcal{G}}
\safemath{\setH}{\mathcal{H}}
\safemath{\setI}{\mathcal{I}}
\safemath{\setJ}{\mathcal{J}}
\safemath{\setK}{\mathcal{K}}
\safemath{\setL}{\mathcal{L}}
\safemath{\setM}{\mathcal{M}}
\safemath{\setN}{\mathcal{N}}
\safemath{\setO}{\mathcal{O}}
\safemath{\setP}{\mathcal{P}}
\safemath{\setQ}{\mathcal{Q}}
\safemath{\setR}{\mathcal{R}}
\safemath{\setS}{\mathcal{S}}
\safemath{\setT}{\mathcal{T}}
\safemath{\setU}{\mathcal{U}}
\safemath{\setV}{\mathcal{V}}
\safemath{\setW}{\mathcal{W}}
\safemath{\setX}{\mathcal{X}}
\safemath{\setY}{\mathcal{Y}}
\safemath{\setZ}{\mathcal{Z}}
\safemath{\emptySet}{\varnothing}

\safemath{\colA}{\mathscr{A}}
\safemath{\colB}{\mathscr{B}}
\safemath{\colC}{\mathscr{C}}
\safemath{\colD}{\mathscr{D}}
\safemath{\colE}{\mathscr{E}}
\safemath{\colF}{\mathscr{F}}
\safemath{\colG}{\mathscr{G}}
\safemath{\colH}{\mathscr{H}}
\safemath{\colI}{\mathscr{I}}
\safemath{\colJ}{\mathscr{J}}
\safemath{\colK}{\mathscr{K}}
\safemath{\colL}{\mathscr{L}}
\safemath{\colM}{\mathscr{M}}
\safemath{\colN}{\mathscr{N}}
\safemath{\colO}{\mathscr{O}}
\safemath{\colP}{\mathscr{P}}
\safemath{\colQ}{\mathscr{Q}}
\safemath{\colR}{\mathscr{R}}
\safemath{\colS}{\mathscr{S}}
\safemath{\colT}{\mathscr{T}}
\safemath{\colU}{\mathscr{U}}
\safemath{\colV}{\mathscr{V}}
\safemath{\colW}{\mathscr{W}}
\safemath{\colX}{\mathscr{X}}
\safemath{\colY}{\mathscr{Y}}
\safemath{\colZ}{\mathscr{Z}}

\safemath{\opA}{\mathbb{A}}
\safemath{\opB}{\mathbb{B}}
\safemath{\opC}{\mathbb{C}}
\safemath{\opD}{\mathbb{D}}
\safemath{\opE}{\mathbb{E}}
\safemath{\opF}{\mathbb{F}}
\safemath{\opG}{\mathbb{G}}
\safemath{\opH}{\mathbb{H}}
\safemath{\opI}{\mathbb{I}}
\safemath{\opJ}{\mathbb{J}}
\safemath{\opK}{\mathbb{K}}
\safemath{\opL}{\mathbb{L}}
\safemath{\opM}{\mathbb{M}}
\safemath{\opN}{\mathbb{N}}
\safemath{\opO}{\mathbb{O}}
\safemath{\opP}{\mathbb{P}}
\safemath{\opQ}{\mathbb{Q}}
\safemath{\opR}{\mathbb{R}}
\safemath{\opS}{\mathbb{S}}
\safemath{\opT}{\mathbb{T}}
\safemath{\opU}{\mathbb{U}}
\safemath{\opV}{\mathbb{V}}
\safemath{\opW}{\mathbb{W}}
\safemath{\opX}{\mathbb{X}}
\safemath{\opY}{\mathbb{Y}}
\safemath{\opZ}{\mathbb{Z}}
\safemath{\opZero}{\mathbb{O}}
\safemath{\identityop}{\opI}

\safemath{\veca}{\bma}
\safemath{\vecb}{\bmb}
\safemath{\vecc}{\bmc}
\safemath{\vecd}{\bmd}
\safemath{\vece}{\bme}
\safemath{\vecf}{\bmf}
\safemath{\vecg}{\bmg}
\safemath{\vech}{\bmh}
\safemath{\veci}{\bmi}
\safemath{\vecj}{\bmj}
\safemath{\veck}{\bmk}
\safemath{\vecl}{\bml}
\safemath{\vecm}{\bmm}
\safemath{\vecn}{\bmn}
\safemath{\veco}{\bmo}
\safemath{\vecp}{\bmp}
\safemath{\vecq}{\bmq}
\safemath{\vecr}{\bmr}
\safemath{\vecs}{\bms}
\safemath{\vect}{\bmt}
\safemath{\vecu}{\bmu}
\safemath{\vecv}{\bmv}
\safemath{\vecw}{\bmw}
\safemath{\vecx}{\bmx}
\safemath{\vecy}{\bmy}
\safemath{\vecz}{\bmz}

\safemath{\veczero}{\bmzero}
\safemath{\vecone}{\bmone}
\safemath{\vecxi}{\bmxi}
\safemath{\veclambda}{\bmlambda}
\safemath{\vecmu}{\bmmu}
\safemath{\vectheta}{\bmtheta}
\safemath{\vecphi}{\bmphi}
\safemath{\vecdelta}{\bmdelta}

\safemath{\matA}{\bA}
\safemath{\matB}{\bB}
\safemath{\matC}{\bC}
\safemath{\matD}{\bD}
\safemath{\matE}{\bE}
\safemath{\matF}{\bF}
\safemath{\matG}{\bG}
\safemath{\matH}{\bH}
\safemath{\matI}{\bI}
\safemath{\matJ}{\bJ}
\safemath{\matK}{\bK}
\safemath{\matL}{\bL}
\safemath{\matM}{\bM}
\safemath{\matN}{\bN}
\safemath{\matO}{\bO}
\safemath{\matP}{\bP}
\safemath{\matQ}{\bQ}
\safemath{\matR}{\bR}
\safemath{\matS}{\bS}
\safemath{\matT}{\bT}
\safemath{\matU}{\bU}
\safemath{\matV}{\bV}
\safemath{\matW}{\bW}
\safemath{\matX}{\bX}
\safemath{\matY}{\bY}
\safemath{\matZ}{\bZ}
\safemath{\matzero}{\bmzero}

\safemath{\matDelta}{\bDelta}
\safemath{\matLambda}{\bLambda}
\safemath{\matPhi}{\bPhi}
\safemath{\matSigma}{\bSigma}
\safemath{\matOmega}{\bOmega}
\safemath{\matTheta}{\bTheta}

\safemath{\matidentity}{\matI}
\safemath{\matone}{\matO}

\safemath{\rnda}{A}
\safemath{\rndb}{B}
\safemath{\rndc}{C}
\safemath{\rndd}{D}
\safemath{\rnde}{E}
\safemath{\rndf}{F}
\safemath{\rndg}{G}
\safemath{\rndh}{H}
\safemath{\rndi}{I}
\safemath{\rndj}{J}
\safemath{\rndk}{K}
\safemath{\rndl}{L}
\safemath{\rndm}{M}
\safemath{\rndn}{N}
\safemath{\rndo}{O}
\safemath{\rndp}{P}
\safemath{\rndq}{Q}
\safemath{\rndr}{R}
\safemath{\rnds}{S}
\safemath{\rndt}{T}
\safemath{\rndu}{U}
\safemath{\rndv}{V}
\safemath{\rndw}{W}
\safemath{\rndx}{X}
\safemath{\rndy}{Y}
\safemath{\rndz}{Z}

\safemath{\rveca}{\bimA}
\safemath{\rvecb}{\bimB}
\safemath{\rvecc}{\bimC}
\safemath{\rvecd}{\bimD}
\safemath{\rvece}{\bimE}
\safemath{\rvecf}{\bimF}
\safemath{\rvecg}{\bimG}
\safemath{\rvech}{\bimH}
\safemath{\rveci}{\bimI}
\safemath{\rvecj}{\bimJ}
\safemath{\rveck}{\bimK}
\safemath{\rvecl}{\bimL}
\safemath{\rvecm}{\bimM}
\safemath{\rvecn}{\bimN}
\safemath{\rveco}{\bomO}
\safemath{\rvecp}{\bimP}
\safemath{\rvecq}{\bimQ}
\safemath{\rvecr}{\bimR}
\safemath{\rvecs}{\bimS}
\safemath{\rvect}{\bimT}
\safemath{\rvecu}{\bimU}
\safemath{\rvecv}{\bimV}
\safemath{\rvecw}{\bimW}
\safemath{\rvecx}{\bimX}
\safemath{\rvecy}{\bimY}
\safemath{\rvecz}{\bimZ}

\safemath{\rvecxi}{\bmxi}
\safemath{\rveclambda}{\bmlambda}
\safemath{\rvecmu}{\bmmu}
\safemath{\rvectheta}{\bmtheta}
\safemath{\rvecphi}{\bmphi}

\safemath{\rmatA}{\bimA}
\safemath{\rmatB}{\bimB}
\safemath{\rmatC}{\bimC}
\safemath{\rmatD}{\bimD}
\safemath{\rmatE}{\bimE}
\safemath{\rmatF}{\bimF}
\safemath{\rmatG}{\bimG}
\safemath{\rmatH}{\bimH}
\safemath{\rmatI}{\bimI}
\safemath{\rmatJ}{\bimJ}
\safemath{\rmatK}{\bimK}
\safemath{\rmatL}{\bimL}
\safemath{\rmatM}{\bimM}
\safemath{\rmatN}{\bimN}
\safemath{\rmatO}{\bimO}
\safemath{\rmatP}{\bimP}
\safemath{\rmatQ}{\bimQ}
\safemath{\rmatR}{\bimR}
\safemath{\rmatS}{\bimS}
\safemath{\rmatT}{\bimT}
\safemath{\rmatU}{\bimU}
\safemath{\rmatV}{\bimV}
\safemath{\rmatW}{\bimW}
\safemath{\rmatX}{\bimX}
\safemath{\rmatY}{\bimY}
\safemath{\rmatZ}{\bimZ}

\safemath{\rmatDelta}{\bimDelta}
\safemath{\rmatLambda}{\bimLambda}
\safemath{\rmatPhi}{\bimPhi}
\safemath{\rmatSigma}{\bimSigma}
\safemath{\rmatOmega}{\bimOmega}
\safemath{\rmatTheta}{\bimTheta}

\usepackage{amssymb}
\usepackage{amsfonts}
\usepackage{mathrsfs}
\usepackage{xspace}
\usepackage{bm}
\usepackage{fancyref}
\usepackage{textcomp}

\usepackage{multirow}
\usepackage{stmaryrd}

\newenvironment{textbmatrix}{	\setlength{\arraycolsep}{2.5pt}%
								\big[\begin{matrix}}{\end{matrix}\big]%
								\raisebox{0.08ex}{\vphantom{M}}}

\def\be{\begin{equation}}
\def\ee{\end{equation}}
\def\een{\nonumber \end{equation}}
\def\mat{\begin{bmatrix}}
\def\emat{\end{bmatrix}}
\def\btm{\begin{textbmatrix}}
\def\etm{\end{textbmatrix}}

\def\ba#1\ea{\begin{align}#1\end{align}}
\def\bas#1\eas{\begin{align*}#1\end{align*}}
\def\bs#1\es{\begin{split}#1\end{split}}
\def\bg#1\eg{\begin{gather}#1\end{gather}}
\def\bml#1\eml{\begin{multline}#1\end{multline}}
\def\bi#1\ei{\begin{itemize}#1\end{itemize}}

\DeclareMathOperator*{\argmin}{arg\;min}		

\safemath{\dirac}{\delta}					
\safemath{\krond}{\dirac}					

\safemath{\upto}{\uparrow}
\safemath{\downto}{\downarrow}
\safemath{\iu}{j}							
\safemath{\ev}{\lambda}						
\safemath{\hilseqspace}{l^{2}}				
\newcommand{\banachfunspace}[1]{\setL^{#1}}	
\safemath{\hilfunspace}{\banachfunspace{2}}	

\safemath{\SNR}{\textit{SNR}} 				
\safemath{\PAR}{\textit{PAR}} 				
\safemath{\No}{N_0}							
\safemath{\Es}{E_s}							
\safemath{\Eb}{E_b}							
\safemath{\EbNo}{\frac{\Eb}{\No}}
\safemath{\EsNo}{\frac{\Es}{\No}}

\DeclareMathOperator{\CHop}{\ensuremath{\opH}} 
\safemath{\tvir}{\rndh_{\CHop}}				
\safemath{\tvtf}{\rndl_{\CHop}}				
\safemath{\spf}{\rnds_{\CHop}}				
\safemath{\bff}{H_{\CHop}}					

\safemath{\ircf}{r_{h}}						
\safemath{\tftvcf}{r_{s}}					
\safemath{\tfcf}{r_{l}}						
\safemath{\bfcf}{r_{H}}						

\safemath{\tcorr}{c_h}						
\safemath{\scf}{c_{s}}						
\safemath{\tfcorr}{c_{l}}					
\safemath{\fcorr}{c_{H}}						

\safemath{\mi}{I}							
\safemath{\capacity}{C}						

\safemath{\normal}{\mathcal{N}}			
\safemath{\jpg}{\mathcal{CN}}			
\safemath{\mchain}{\leftrightarrow}		

\safemath{\dB}{\,\mathrm{dB}}
\safemath{\dBm}{\,\mathrm{dBm}}
\safemath{\Hz}{\,\mathrm{Hz}}
\safemath{\kHz}{\,\mathrm{kHz}}
\safemath{\MHz}{\,\mathrm{MHz}}
\safemath{\GHz}{\,\mathrm{GHz}}
\safemath{\s}{\,\mathrm{s}}
\safemath{\ms}{\,\mathrm{ms}}
\safemath{\mus}{\,\mathrm{\text{\textmu}s}}
\safemath{\ns}{\,\mathrm{ns}}
\safemath{\ps}{\,\mathrm{ps}}
\safemath{\meter}{\,\mathrm{m}}
\safemath{\mm}{\,\mathrm{mm}}
\safemath{\cm}{\,\mathrm{cm}}
\safemath{\m}{\,\mathrm{m}}
\safemath{\W}{\,\mathrm{W}}
\safemath{\mW}{\, \mathrm{mW}}
\safemath{\J}{\,\mathrm{J}}
\safemath{\K}{\,\mathrm{K}}
\safemath{\bit}{\,\mathrm{bit}}
\safemath{\nat}{\,\mathrm{nat}}

\safemath{\define}{\triangleq}			

\safemath{\equivalent}{\sim}
\safemath{\distas}{\sim}					
\safemath{\sdiff}{\Delta}				

\safemath{\reals}{\mathbb{R}}
\safemath{\positivereals}{\reals_{+}}
\safemath{\integers}{\mathbb{Z}}
\safemath{\posint}{\integers_{+}}
\safemath{\naturals}{\mathbb{N}}
\safemath{\posnaturals}{\naturals_{+}}
\safemath{\complexset}{\mathbb{C}}
\safemath{\rationals}{\mathbb{Q}}

\newcommand*{\fancyrefapplabelprefix}{app}		
\newcommand*{\fancyrefthmlabelprefix}{thm}		
\newcommand*{\fancyreflemlabelprefix}{lem}		
\newcommand*{\fancyrefcorlabelprefix}{cor}		
\newcommand*{\fancyrefdeflabelprefix}{def}		
\newcommand*{\fancyrefproplabelprefix}{prop}		
\newcommand*{\fancyrefexmpllabelprefix}{exmpl}
\newcommand*{\fancyrefalglabelprefix}{alg}		
\newcommand*{\fancyreftbllabelprefix}{tbl}		

\frefformat{vario}{\fancyrefseclabelprefix}{Sec.~#1}
\frefformat{vario}{\fancyrefthmlabelprefix}{Thm.~#1}
\frefformat{vario}{\fancyreftbllabelprefix}{Tbl.~#1}
\frefformat{vario}{\fancyreflemlabelprefix}{Lem.~#1}
\frefformat{vario}{\fancyrefcorlabelprefix}{Cor.~#1}
\frefformat{vario}{\fancyrefdeflabelprefix}{Def.~#1}
\frefformat{vario}{\fancyreffiglabelprefix}{Fig.~#1}
\frefformat{vario}{\fancyrefapplabelprefix}{App.~#1}
\frefformat{vario}{\fancyrefeqlabelprefix}{(#1)}
\frefformat{vario}{\fancyrefproplabelprefix}{Prop.~#1}
\frefformat{vario}{\fancyrefexmpllabelprefix}{Ex.~#1}
\frefformat{vario}{\fancyrefalglabelprefix}{Alg.~#1}

\safemath{\dictab}{[\,\dicta\,\,\dictb\,]}

\safemath{\ysig}{\bmy}
\safemath{\ysighat}{\hat{\ysig}}
\safemath{\ysigdim}{M}
\safemath{\xsig}{\bmx}
\safemath{\xsigdim}{N}
\safemath{\nx}{n_x}
\safemath{\zsig}{\bmz}
\safemath{\zsigdim}{\ysigdim}
\safemath{\rsig}{\bmr}
\safemath{\Adict}{\bA}
\safemath{\Adicttilde}{\widetilde{\Adict}}
\safemath{\Adictdim}{\outputdim\times\xsigdim}
\safemath{\avec}{\bma}
\safemath{\avectilde}{\tilde{\avec}}
\safemath{\Bdict}{\bB}
\safemath{\Bdicttilde}{\widetilde{\Bdict}}
\safemath{\Cdict}{\bC}
\safemath{\cvec}{\bmc}
\safemath{\Ddict}{\bD}
\safemath{\Ddictdim}{\ysigdim\times\xsigdim}
\safemath{\dvec}{\bmd}
\safemath{\Ddicttilde}{\widetilde{\bD}}
\safemath{\Bonb}{\bB}
\safemath{\bvec}{\bmb}
\safemath{\Bonbdim}{\ysigdim\times\ysigdim}
\safemath{\noise}{\bmn}
\safemath{\noisedim}{\ysigim}
\safemath{\err}{\bme}
\safemath{\errdim}{\ysigdim}
\safemath{\errset}{\setE}
\safemath{\nerr}{n_e}
\safemath{\delop}{\bP_\errset}
\safemath{\delopc}{\bP_{{\errset}^c}}

\safemath{\cplxi}{\imath}
\safemath{\cplxj}{\jmath}

\safemath{\dict}{\matD}
\safemath{\inputdim}{N}		
\safemath{\outputdim}{M}		
\safemath{\sparsity}{S}	
\safemath{\inputdimA}{{N_a}}	
\safemath{\inputdimB}{{N_b}}	
\safemath{\elemA}{{n_a}}	
\safemath{\elemB}{{n_b}}	
\safemath{\resA}{\matR_a}	
\safemath{\resB}{\matR_b}	
\safemath{\subD}{\matS} 
\safemath{\subA}{\matS_a} 
\safemath{\subB}{\matS_b} 
\safemath{\dicta}{\matA} 	
\safemath{\dictb}{\matB} 	
\safemath{\hollowS}{H}
\safemath{\hollowA}{H_a}
\safemath{\hollowB}{H_b}
\safemath{\cross}{Z}
\safemath{\coh}{\mu_d}			
\safemath{\coha}{\mu_a}			
\safemath{\cohb}{\mu_b}			
\safemath{\mubs}{\nu}	
\safemath{\cohm}{\mu_m} 
\safemath{\dictset}{\setD}	
\safemath{\dictsetp}{\dictset(\coh,\coha,\cohb)}	
\safemath{\dictsetgen}{\dictset_\text{gen}}
\safemath{\dictsetgenp}{\dictsetgen(\coh)}
\safemath{\dictsetonb}{\dictset_\text{onb}}
\safemath{\dictsetonbp}{\dictsetonb(\coh)}

\safemath{\leftside}{U}
\safemath{\rightsideA}{R_a}
\safemath{\rightsideB}{R_b}

\safemath{\indexS}{\setI_S} 

\safemath{\na}{n_a}			
\safemath{\nb}{n_b}			
\safemath{\coeffa}{p_i}	
\safemath{\coeffb}{q_j}	
\safemath{\seta}{\setP}		
\safemath{\setb}{\setQ}     
\safemath{\setw}{\setW}	
\safemath{\setz}{\setZ}	
\safemath{\cola}{\veca}		
\safemath{\colb}{\vecb}		
\safemath{\cold}{\vecd}		
\safemath{\inputvec}{\vecx} 	
\safemath{\error}{\vece}	
\safemath{\noiseout}{\vecz} 	
\safemath{\inputvecel}{x}
\safemath{\inputveca}{\vecx_a}
\safemath{\inputvecb}{\vecx_b}
\safemath{\outputvec}{\vecy}	
\safemath{\lambdamin}{\lambda_{\mathrm{min}}}

\safemath{\elltwo}{\ell_2}
\safemath{\ellone}{\ell_1}
\safemath{\ellzero}{\ell_0}
\safemath{\ellinf}{\ell_\infty}
\safemath{\ellinftilde}{\ell_{\widetilde\infty}}
\safemath{\licard}{Z(\coh,\coha,\cohb)}
\safemath{\xsol}{\hat{x}}
\safemath{\xbord}{x_b}		
\safemath{\xstat}{x_s}		
\safemath{\xstatLone}{\tilde{x}_s}
\safemath{\order}{\mathcal{O}} 
\safemath{\scales}{\Theta} 
\safemath{\ones}{\mathbf{1}} 
\safemath{\zeroes}{\mathbf{0}} 
\safemath{\thlone}{\kappa(\coh,\cohb)} 
\safemath{\constoneA}{\delta} 
\safemath{\constoneB}{\epsilon} 
\safemath{\nlarge}{L}				   
\safemath{\sumlarge}{S_\nlarge}
\safemath{\maxlarger}{P_\nlarge}	   
\safemath{\Pzero}{\textrm{P0}}	
\safemath{\Pone}{\textrm{P1}}
\safemath{\vecfir}{\vecw}			 
\safemath{\vecsec}{\vecz}
\safemath{\elvecfir}{w}              
\safemath{\elvecsec}{z}				 
\safemath{\nlargefir}{n}
\safemath{\normout}{\gamma}
\safemath{\auxfun}{h}
\safemath{\supp}{\textrm{supp}}

\safemath{\indexa}{\ell}
\safemath{\indexb}{r}
\safemath{\indexc}{i}
\safemath{\indexd}{j}

\safemath{\project}{P}

\newcommand*{\fancyrefremarklabelprefix}{remark}
\frefformat{vario}{\fancyrefremarklabelprefix}{Remark~#1}

\usepackage{color}
\def\MSE{\mathrm{MSE}}
\newcommand{\xap}[1]{\text{\underbar{$\vecx$}}^{(#1)}}
\newcommand{\tildexap}[1]{\text{\underbar{$\tilde\vecx$}}^{(#1)}}
\def\loss{\mathfrak{L}}

\def\combinedloss{\loss_{\textnormal{m}}}
\def\tripletloss{\loss_{\textnormal{t}}}
\def\biloss{\loss_{\textnormal{bi}}}
\def\boxloss{\loss_{\textnormal{box}}}
\def\lambdat{\lambda_{\textnormal{t}}}
\def\lambdabi{\lambda_{\textnormal{bi}}}
\def\lambdabox{\lambda_{\textnormal{box}}}
\def\Tc{T_{\textnormal{c}}}
\def\Mt{M_{\textnormal{t}}}
\def\Mb{M_{\textnormal{b}}}
\def\Mp{M_{\textnormal{p}}}
\def\xmin{x_{\textnormal{min}}}
\def\xmax{x_{\textnormal{max}}}
\def\ymin{y_{\textnormal{min}}}
\def\ymax{y_{\textnormal{max}}}
\def\Pthr{P_{\textnormal{thr}}}
 
\newtheoremstyle{revcom}
{\baselineskip} {0.5\baselineskip}		
{\itshape}								
{0 pt}									
{\bfseries}{:}							
{ }										
{}										

\theoremstyle{revcom}
\newtheorem{commentrev}{Comment}
\newtheorem*{commentrev*}{Comment}

\newtheoremstyle{feedbackstyle}
  {}
  {}
  {\itshape}
  {}
  {\bfseries}
  {}
  { }
  {\thmname{#1}\thmnote{#3}}
\makeatother

\theoremstyle{feedbackstyle}
\newtheorem*{commentfeed}{Feedback:}



\title{Channel Charting in Real-World Coordinates\\ with Distributed MIMO\author{\IEEEauthorblockN{Sueda Taner, Victoria Palhares, and Christoph Studer}\thanks{S. Taner, V. Palhares, and C. Studer are with the Department of Information Technology and Electrical Engineering, ETH Zurich, Switzerland (taners@iis.ee.ethz.ch, vmenescal@ethz.ch, and studer@ethz.ch).}%
\thanks{The work of ST, VP, and CS was supported in part by an ETH Research Grant, by the Swiss National Science Foundation (SNSF) grant 200021\_207314, and by CHIST-ERA grant for the project CHASER (CHIST-ERA-22-WAI-01) through the SNSF grant 20CH21\_218704.}
\thanks{The authors thank Remcom for providing a license for the Wireless InSite ray-tracing software. The authors also thank Olav Tirkkonen and Maxime Guillaud for discussions on channel charting in real-world coordinates.}%
\thanks{A preliminary version of this work was presented at the 2023 IEEE GLOBECOM~\cite{taner2023globecom}. In \cite{taner2023globecom}, the bilateration loss was proposed for channel charting in real-world coordinates for line-of-sight (LoS) scenarios. Here, we extend our approach to mixed LoS and non-LoS scenarios by modifying the bilateration loss and by introducing a novel LoS bounding-box loss. In addition, we include experimental results with measured channel vectors from~\cite{dataset-dichasus-cf0x}.}%
}}

\begin{document}

	
\maketitle


\begin{abstract}
Channel charting is an emerging self-supervised method that maps channel-state information (CSI) to a low-dimensional latent space (the channel chart) that represents pseudo-positions of user equipments (UEs).
While channel charts preserve local geometry, i.e., nearby UEs are nearby in the channel chart (and vice versa), the pseudo-positions are in arbitrary coordinates and global geometry is typically not preserved.
In order to embed channel charts in real-world coordinates, we first propose a bilateration loss for distributed multiple-input multiple-output (D-MIMO) wireless systems in which only the access point (AP) positions are known. 
The idea behind this loss is to compare the received power at pairs of APs to determine whether a UE should be placed closer to one AP or the other in the channel chart. 
We then propose a line-of-sight (LoS) bounding-box loss that places the UE in a predefined LoS area of each AP that is estimated to have a LoS path to the UE.
We demonstrate the efficacy of combining both of these loss functions with neural-network-based channel charting using ray-tracing-based and measurement-based channel vectors.
Our proposed approach outperforms several baselines and maintains the self-supervised nature of channel charting as it neither relies on geometrical propagation models nor on any ground-truth UE position information.
\end{abstract}




\section{Introduction} 
\label{sec:intro}

Channel charting is a self-supervised method that extracts user equipment (UE) pseudo-position solely by processing estimated channel-state information (CSI) at infrastructure basestations (BSs) or access points (APs)~\cite{studer18cc}. 
The key idea is to apply dimensionality reduction to a large database of CSI features that represent large-scale fading properties of the wireless channel, such as time-of-arrival (ToA), angle-of-arrival (AoA), power-delay profile, etc.
The low-dimensional latent space resulting from dimensionality reduction is the so-called  \emph{channel chart}, which is tied to UE position. More specifically, UEs nearby in real space are nearby in the channel chart and vice versa. 
The learned channel chart can be used to assist a wide range of applications in wireless systems that rely on UE position information, such as pilot allocation \cite{ribeiro22ojcs}, beam management\mbox{\cite{luclemag22,kazemi22vtc}}, channel capacity prediction \cite{kallehague23}, and many more; see~\cite{ferrand2023wireless} for an overview.

Although channel charts preserve the local geometry of UE positions, i.e., nearby UEs correspond to points that are nearby in the channel chart (and the other way around), global geometry is typically distorted, e.g., scaled, rotated, and warped. Furthermore, the learned channel charts are represented in arbitrary coordinates, which is a direct consequence of self-supervised learning. 
Multipoint channel charting techniques, which process CSI acquired at multiple distributed multi-antenna APs, have been shown to improve global geometry\mbox{\cite{deng18,euchner22,agostini22federated}}. 
Nonetheless, the resulting channel charts remain to be pseudo-positions with no ties to real-world coordinates.
In this paper, we resolve exactly this limitation: 
We aim to learn channel charts that are interpretable in terms of real-world coordinates.
{Learning such a channel chart in real-world coordinates has the following benefits and example use cases:}
\begin{itemize}
\item {Efficient handovers: Mapping CSI to real-world coordinates would enable one to predict UE trajectories (considering the UE speed and direction), enabling successful handover to the best next cell. 
}
\item {Improved beamforming: Having an estimate of the UE positions would improve beam alignment for sending strong signals into desired directions.
}
\item {Dynamic resource allocation: 
Having an estimate of UE positions would allow improved pilot allocation and user scheduling; 
for example, the BS could allocate different pilots and time slots for UEs that are close to each other. 
}
\item {Smarter network deployment planning: Building a position map over time would reveal traffic patterns and coverage gaps, enabling better placement of APs. 
}
\end{itemize} 
{Such use cases can be improved by channel charting in real-world coordinates because near as well as far distances in the channel chart now have a physical meaning, in contrast to traditional channel charting in which only nearby distance information is~reliable.}

\subsection{Contributions} 
\label{sec:contributions}
We propose novel methods that learn neural-network-based channel charting functions in a distributed multiple-input multiple-output (D-MIMO) scenario to generate channel charts in real-world coordinates.
Our key contributions are {summarized} as follows:
\begin{itemize}
\item {
We extend the bilateration loss proposed in~\cite{taner2023globecom} for line-of-sight (LoS) scenarios to accommodate scenarios with non-LoS  propagation by excluding non-LoS APs from the loss calculation;
here, we determine the non-LoS APs based on a power threshold.
}
\item {
We propose a novel LoS bounding-box loss, which places the UE in the LoS area of the APs whose receive power is higher than the above-mentioned power threshold. }
\item {
We propose to combine the bilateration and the LoS bounding-box losses with the timestamp-based triplet loss from~\cite{ferrand2021} for improved channel chart quality.}
\item {
We demonstrate the efficacy of our method in one indoor and one outdoor scenario based on channel vectors from a commercial ray-tracer \cite{remcom} and another indoor scenario based on real-world measurements~\cite{dataset-dichasus-cf0x}. }
\end{itemize}
We emphasize that our proposed methods are weakly supervised as they only require (i) knowledge of the AP positions and (ii) approximate knowledge of the LoS areas surrounding each AP---\emph{no} geometric models, accurate AP synchronization, or labeled CSI samples with UE ground-truth positions are required.

\subsection{Relevant Prior Art}

The literature describes a plethora of methods for radio-based positioning 
\cite{WEN201921survey,liu2007indoorsurvey,localizationsurvey18,liu2019indoorsurvey}.
While our main goal is not accurate positioning, we briefly review these methods for completeness.
Geometric model-based positioning methods, e.g., methods that rely on ToA and/or AoA (see \cite{widdison24review} and the references therein), can achieve precise positioning in environments with LoS connectivity. 
These methods, however, rely on accurately synchronized APs, and their performance degrades in non-LoS and multipath scenarios.
An alternative radio-based positioning technique which does not require AP synchronization is fingerprinting~\cite{he16survey,wang2017csi,lundpaper,arnold2019novel,ferrand2020globecom,stahlke22vtc,gonultas22twc,foliadis22icc,li23vtc,tian23tmlcn}.
CSI fingerprinting methods rely on the offline collection of a large CSI database where the ground-truth UE position corresponding to each CSI sample must be known.
However, this approach is costly since setting up a ground-truth reference system is both cumbersome and expensive; moreover, the ground-truth labeled CSI measurement campaign needs to be repeated whenever there is a significant change in the physical environment.

In order to reduce the number of ground-truth labeled CSI samples required by fingerprinting-based methods, semi-supervised learning methods have been proposed~\cite{penghzi19spawc,lei19siamese,deng21networkside,zhang21globecom,deng21tsne}.
This strain of methods assumes that the ground-truth UE positions are known for a (small) subset of the training dataset, and incorporates these known positions into learning a channel charting function.
{Addressing the ground-truth label acquisition issue from a different point of view, reference \cite{ermolov23globecom}  utilizes pseudolabels from an inertial measurement unit (IMU) to overcome the dependency on hard-to-acquire precise position labels. 
}
In contrast to such approaches, which still require a ground-truth reference positioning system, our method does not require ground-truth position information for the CSI samples.

More recently, another strain of methods that incorporates real-world positions directly into channel charting has been proposed in~\cite{karmanov21wicluster,euchner23asilomar, esrafilian24arxiv}. 
Reference \cite{karmanov21wicluster} relies on several zone-labeled CSI samples to predict a zone for every CSI sample; this zone information is then used during weakly-supervised training. 
References \cite{euchner23asilomar, esrafilian24arxiv} propose loss functions that leverage estimated ToA or AoA to assist channel charting.
As we have mentioned before, these approaches require accurately synchronized APs.
In contrast, we require neither zone-labeled CSI samples nor accurate AP synchronization.

Several methods have been proposed that first create a channel chart in a self-supervised manner, then use an affine transform that maps the channel chart to real-world positions~\cite{pihljasalo20,stahlke23,stahlke2023velocity,stephan2023adp,euchner23asilomar}.
Various methods have been proposed to learn such affine transforms:
The method in \cite{pihljasalo20} estimates the AP positions in the channel chart and maps these positions to the real-world positions of the APs.
The method in \cite{stahlke2023velocity} assumes that a map of the environment is known a priori and aims to match the channel chart to this environment map.
The methods in \cite{stahlke23,stephan2023adp} use several ground-truth UE positions,\footnote{We make a distinction between methods that use ground-truth positions \emph{during} neural-network training~\cite{he16survey,wang2017csi,lundpaper,arnold2019novel,ferrand2020globecom,stahlke22vtc,gonultas22twc,foliadis22icc,li23vtc,tian23tmlcn,penghzi19spawc,lei19siamese,deng21networkside,zhang21globecom,deng21tsne} and methods that use ground-truth positions \emph{after} training \cite{stahlke23,stephan2023adp}. We only use the terms ``supervised'' and ``semi-supervised'' to refer to the first kind of methods. 
} and the method in \cite{euchner23asilomar} uses UE position estimates obtained through geometric model-based positioning techniques.
In contrast to such transformation-based methods, which would fail when channel charts have non-affine distortions of the real-world positions and may require labeled CSI samples, we aim to directly learn channel charts that are embedded in real-world coordinates in a weakly-supervised manner.

\subsection{Notation}

Column vectors and matrices are denoted by lowercase and uppercase boldface letters, respectively; sets are denoted by uppercase calligraphic letters.
The column-wise vectorization of $\bA$ is denoted by $\mathrm{vec}(\bA)$.    
For a vector~$\bma$, 
the Euclidean norm is $\|\bma\|$ and the entry-wise absolute value is $|\bma|$; for a matrix $\bA$, the Frobenius norm is $\|\bA\|_F$. 
The operator $(x)^+=\max\{x,0\}$ is the rectified linear unit (ReLU).
The indicator function is denoted by $\mathds{1}_{\{c\}}$, which is one if the condition $c$ is met and zero otherwise.

\subsection{Paper Outline}
The rest of the paper is organized as follows. \fref{sec:background} introduces the basics of channel charting.
\fref{sec:ccinreal} proposes novel loss functions that embed channel charts in real-world coordinates.
\fref{sec:eval_setup} details the proposed methods, baselines, and performance metrics.
\fref{sec:results} presents simulation results for three different scenarios.
\fref{sec:conclusions} concludes the paper.



\section{Channel Charting Basics}
\label{sec:background}

We now briefly outline the basics of channel charting, detail the system model, and discuss parametric channel charting with neural networks using the triplet-based learning approach~\cite{ferrand2021}, which is part of the method we propose in \fref{sec:ccinreal}.

\subsection{Operating Principle }
Channel charting typically operates in two phases~\cite{studer18cc}. In the first phase, CSI from a large number of different UE positions is acquired and CSI features that capture large-scale fading properties of the wireless channel are stored in a database. By applying parametric dimensionality reduction~\cite{vandermaaten2009dimensionality} to the CSI feature database, one then learns a channel charting function (e.g., implemented as a neural network), which maps CSI features to a low-dimensional representation: the channel chart.
Channel charts have the useful property that the UEs transmitting from nearby positions are also placed nearby in the channel chart and vice versa. 
We reiterate that channel-chart learning is self-supervised, meaning that no ground-truth information about the UEs' position is required.
In the second phase, the channel charting function is used to map new CSI features to points in the channel chart, which represent the transmitting UEs' pseudo-positions.

\subsection{System Model}
\label{sec:system_model}

We consider a D-MIMO wireless system in which one or multiple single-antenna UEs transmit pilots to $A$ distributed APs with $M_R$ antennas each, leading to $B=A M_R$ receive antennas in total. 
We consider orthogonal frequency division multiplexing (OFDM) transmission with $W$ occupied subcarriers.
We assume that the AP $a\in\setA=\{1,\dots,A\}$ is at position $\xap{a}\in\reals^3$ in physical space, and that these positions are known.
Moreover, we assume approximate knowledge of {the LoS area of each AP represented by the minimum and maximum $x$ and $y$ coordinates of the room (or open space) containing the AP}\footnote{
It is a reasonable assumption to know the AP positions and their LoS areas (e.g., the {size and shape of the} room the AP is in), as the APs are typically not placed arbitrarily but according to a carefully-crafted deployment plan. \label{footnote_knownlosbox}}, 
denoted as $\xmin^{(a)}$, $\xmax^{(a)}$, $\ymin^{(a)}$, $\ymax^{(a)} \in\opR$, {which define a rectangular bounding box that encloses the LoS area}\footnote{{
We decided to use rectangular bounding boxes because (i) rooms are often rectangular, making our assumption practical, and (ii) rectangular bounding boxes provide simple mathematical expressions that are easier to implement when training neural networks compared to arbitrary shapes (e.g., polygons); exploring arbitrary shapes is left for future work.}\label{footnote_rectangle}}. 
We denote the LoS bounding box for AP $a$ by 
\begin{align}
\setB^{(a)} = \big\{(x,y)\in\opR^2: x\in [x_{\text{min}}^{(a)} , x_{\text{max}}^{(a)} ], y \in [y_{\text{min}}^{(a)} , y_{\text{max}}^{(a)} ] \big\}. \label{eq:boundingboxset}    
\end{align}

Suppose that we have $N$ pilot transmissions from one UE at (unknown) positions $\bmx^{(n)}\in\opR^3$ at timestamps $t_n$ for $n\in\setN=\{1,\dots,N\}$.
The $n$th transmission from UE position~$\bmx^{(n)}$ enables the AP $a$ to estimate the associated CSI vector $\bmh^{(n,a)}_w\in\opC^{M_R}$ at timestamp~$t_n$ and subcarrier $w$.
By stacking the channel vectors $\{\bmh^{(n,a)}_w\}_{a=1}^A$ from all APs, we can construct a CSI vector $\bmh^{(n)}_w\in\opC^B$ that contains CSI for all $B$ receive antennas at subcarrier $w$. 
It is important to note that we do not require the $A$ APs to be perfectly synchronized while acquiring CSI; 
the only requirement is that the CSI estimated at each AP belongs to the same UE transmitting from approximately the same position $\vecx^{(n)}$ at approximately the same timestamp $t_n$.
Finally, we concatenate the channel vectors from all subcarriers to construct a $B\times W$ CSI matrix associated with the UE at timestamp~$t_n$ as $\bH^{(n)}=[\bmh_1^{(n)},\dots,\bmh_W^{(n)}]$.
The entire CSI database is given by the set of matrices $\{\bH^{(n)}\}_{n\in \setN}$.

\subsection{CSI Feature Extraction}
\label{sec:feature_extraction}

In order to be agnostic to small-scale fading effects and resilient to system and hardware impairments (e.g., phase offsets between APs), we extract large-scale fading properties of the wireless channel by transforming the estimated CSI matrices into CSI features~\cite{studer18cc,ferrand2021,gonultas22twc}.
In what follows, we utilize the CSI features from~\cite{lundpaper,lei19siamese}. 
First, we transform frequency-domain CSI into the (approximate) delay domain by applying an inverse discrete Fourier transform over the~$W$ occupied subcarriers. 
Since most of the received power should be concentrated on the first few taps, we truncate the delay-domain CSI matrix by taking its first $C\ll W$ columns; we denote the truncated delay-domain CSI matrix by $\hat\bH^{(n)}\in\opC^{B\times C}$.
We vectorize this matrix to $\hat\vech^{(n)} = \mathrm{vec}(\hat \bH^{(n)}) \in \opC^{D'}$, where $D'=BC$.
Finally, we compute the unit-norm CSI feature vector $\vecf^{(n)}\in \opR^{D'}$ as
\begin{align}
\vecf^{(n)} = \frac{|\hat\vech^{(n)}|}{\| \hat\vech^{(n)}\|}, 
\end{align} 
which ignores (i) phase shifts (e.g., stemming from APs that are not accurately synchronized) by taking the entry-wise absolute value and (ii) the UEs' transmit power (e.g., as the UEs can set their own transmit power) by normalizing the resulting vector. 
The entire CSI feature dataset is then given by the set of CSI feature vectors~$\{\vecf^{(n)}\}_{n\in \setN}$.

\subsection{Channel Charting with Neural Networks }

With the CSI feature database, one can then learn a channel-charting function $g_{\boldsymbol\theta}:\opR^{D'}\to\opR^{D}$, which maps a CSI feature vector $\vecf^{(n)}$ to a $D$-dimensional pseudo-position in the channel chart as follows: $\hat\vecx^{(n)} = g_{\boldsymbol\theta}(\vecf^{(n)})$.
In this paper, we implement the function~$g_{\boldsymbol\theta}$ as a neural network that is parametrized by the vector~$\boldsymbol\theta$ which includes all of the weights and biases. 
While real-world positions are in three-dimensional space with coordinates $(x,y,z)$, the channel chart can be embedded in fewer dimensions if, for example, the $z$-coordinate is fixed across the AP and UE positions. Hence, assuming that $D\leq 3$, we denote the truncated vector consisting of the first $D$ entries of $\vecx^{(n)}$ and $\xap{a}$ as~$\tilde\vecx^{(n)}\in\opR^D$ and~$\tildexap{a}\in\opR^D$, respectively.

The literature describes a variety of  methods to learn neural-network-based channel charting functions, such as autoencoders~\cite{studer18cc,penghzi19spawc}, 
Siamese neural networks \cite{lei19siamese}, and neural networks trained with a timestamp-based triplet-loss~\cite{ferrand2021,rappaport2021,yassine22}. 
In what follows, we focus on the timestamp-based triplet-loss approach from~\cite{ferrand2021}.

\subsection{Triplet-Based Channel Charting }
\label{sec:triplet_cc}

Assuming that the timestamps associated with all CSI features are available, reference~\cite{ferrand2021} proposes to use this side information when comparing pairwise distances in latent space.
To this end, one defines a set of triplets from the set of sample indices $\setN$ as follows:
\begin{align}
\setT = \{(n,c,f) \in \setN^3: 0 < |t_n-t_c| \leq \Tc <|t_n-t_f|\},
\end{align}
where $\Tc>0$ is a coherence-time parameter that categorizes the CSI features as close or far in time.
If $t_n$ is closer to $t_c$ than $t_f$, then we would expect the Euclidean distance between the UE positions at timestamp~$t_n$ and~$t_c$ to be smaller than that of $t_f$.
This property can be expressed with the following triplet loss~\cite{ferrand2021}: 
\begin{align} \label{eq:loss_t}
\loss_{\text{t}} = \frac{1}{|\setT|} \! \sum_{(n,c,f)\in\setT} \!\! \big(& \|g_{\boldsymbol\theta}(\vecf^{(n)}) - g_{\boldsymbol\theta}(\vecf^{(c)})\| \notag\\[-0.2cm]
& - \|g_{\boldsymbol\theta}(\vecf^{(n)}) - g_{\boldsymbol\theta}(\vecf^{(f)})\| + \Mt\big)^+.
\end{align}
Here, the margin parameter~$\Mt\geq0$ enforces $g_{\boldsymbol\theta}(\vecf^{(n)})$ to be at least $\Mt$ closer to $g_{\boldsymbol\theta}(\vecf^{(c)})$ than to $g_{\boldsymbol\theta}(\vecf^{(f)})$.

As demonstrated in~\cite{ferrand2021}, one can train a neural network that implements the channel charting function $g_{\boldsymbol\theta}$ by minimizing the loss $\loss_{\text{t}}$; this approach is self-supervised as the loss only utilizes the training dataset without any additional ground-truth labels, such as UE positions.  
We reiterate that the coordinate system of the resulting channel chart is arbitrary (e.g., scaled, rotated, and globally warped).
The methods we propose next address exactly this limitation of existing channel charting pipelines.



\section{Channel Charting in Real-World Coordinates}
\label{sec:ccinreal}

We now propose two loss functions that enable one to directly learn channel charts in real-world coordinates.
Our first loss function uses the known AP positions in a D-MIMO scenario. Our second loss relies on known bounding boxes for the LoS area surrounding each AP. We reiterate that \emph{no} geometric models, accurate time synchronization among APs, or  CSI samples labeled with UE positions are required.

Both of the loss functions we propose rely on LoS APs.
Hence, we have the following simple procedure to estimate whether an AP has an LoS or non-LoS channel with the $n$th UE position:
Let $\hat\bH^{(n,a)} \in \opC^{M_R\times C}$ denote the submatrix of $\hat\bH^{(n)}$ which consists of the $M_R$ rows corresponding to AP~$a$.
We compute the receive power associated with the UE and the $a$th AP as $P^{(n,a)} = 20\log_{10} ( \| \hat\bH^{(n,a)}\|_F )$, $a\in\setA$.
By plotting the values of $P^{(n,a)}$ with respect to $t_n$, we can observe where the power has an instantaneous, large drop, and manually set a power threshold $\Pthr$; 
we provide concrete examples for choosing this threshold in~\fref{sec:indoor}.
We classify the APs whose receive power is above this threshold as ``in LoS,'' and the remaining ones as ``not in LoS.''
We construct the set of (estimated) LoS APs for each UE position as follows\footnote{{We employ a power-thresholding approach for LoS/non-LoS classification for the sake of simplicity. We refer to \cite{kirmaz21pimrc,brasseler24nonlos,lopez23polarization} and the references therein for more advanced LoS/non-LoS classification methods that come at the cost of higher computational complexity and possibly more hyperparameters.}\label{footnote_los}}:
\begin{align}
\tilde\setA^{(n)} = \{ a\in\setA : P^{(n,a)} > \Pthr \} . \label{eq:tildeA}
\end{align}
We utilize $\tilde\setA^{(n)}$ in our loss functions detailed next.

\subsection{Bilateration Loss}
\label{sec:loss_bi}

Assuming that a UE's pilot signal is received at multiple APs in the considered D-MIMO scenario, our bilateration loss aims to utilize the known AP positions and place the UE closer to one AP than another.
Here, our CSI-based clue for the real-world position of the UE is the relative power of the CSI matrix for each LoS AP:
As one would expect, an AP with higher relative receive power tends to be closer to the UE under LoS conditions, whereas the power would give no clue on the distance under non-LoS conditions. 
Hence, we only utilize the APs from $\tilde\setA^{(n)}$ in \fref{eq:tildeA} for the bilateration loss. 
Specifically, we define the following set of AP pairs:
\begin{align} \label{eq:setP}
\setP^{(n)} \!=\! \big\{& (a_c,a_f) \in (\tilde\setA^{(n)})^2  : P^{(n,a_c)} > P^{(n,a_f)} + \Mp \big\}.
\end{align}
The set $\setP^{(n)}$ generates pairs of AP indices $(a_c,a_f)$, where the two selected APs act as close and far reference points for the $n$th UE position. 
Here, the margin parameter $\Mp\geq 0$ enforces the set $\setP^{(n)}$ to only include AP pairs whose channel powers differ by at least $\Mp$.
The assumption that the UE $n$ should be closer to AP $a_c$ than AP $a_f$ as determined by the set $\setP^{(n)}$ in \fref{eq:setP} leads to the following \textit{bilateration loss}:
\begin{align} \label{eq:loss_bi}
\biloss=\, &  { \frac{1}{\sum_{n\in\setN} |\setP^{(n)}|} }\sum_{n\in\setN} \sum_{(a_c,a_f)\in \setP^{(n)}} \notag \\
& \big( \|g_{\boldsymbol\theta}(\vecf^{(n)}) \!-\! \tildexap{a_c}\|  \!-\! \|g_{\boldsymbol\theta}(\vecf^{(n)}) \!-\! \tildexap{a_f}\|+ \Mb\big)^+\!\!.
\end{align}
Here, the margin $\Mb\geq0$ enforces $g_\theta(\vecf^{(n)})$ to be at least $\Mb$ closer to AP $a_c$ than AP $a_f$.
We emphasize that the loss $\biloss$ requires no assumption of how far each AP should be from the UE based on their channel power; we merely deduce relative distances to the two APs. Moreover,
while the underlying assumption on the power-distance relation between UE and APs may not always hold, we seek no such guarantee---this claim is supported with a concrete example in~\fref{sec:outdoor_power}.

One can train a neural network that implements the channel charting function $g_{\boldsymbol\theta}$ by minimizing the loss $\biloss$ {in \fref{eq:loss_bi}; 
this approach is weakly-supervised as the loss incorporates only partial information on the ground-truth labels (i.e., only position information of the APs is used).

We note that a loss based on received power was also proposed in~\cite{karmanov21wicluster}; this loss, however, compares the power of two CSI vectors corresponding to the channel between one AP and the UE at two timestamps $t_i$ and $t_k$  
in order to deduce whether $\vecx^{(i)}$ or $\vecx^{(k)}$ is closer to the AP. 
This loss does not leverage the fact that multiple APs receive the UE's pilot signal simultaneously.
In contrast, our bilateration loss in \fref{eq:loss_bi} takes each CSI feature into account individually and compares the receive power between a pair of APs; this means that our loss does not rely on timestamps.
Thus, our bilateration loss could also be used in scenarios in which timestamps are irrelevant, e.g., a multiuser scenario with stationary UEs.

{We note that the proposed bilateration loss suffers from the following limitations. First, it can only be utilized for UE positions that have an LoS path to at least one pair of APs. Second, it may not be sufficient to accurately estimate the $n$th UE position if the set $\setP^{(n)}$ does not include multiple AP pairs.
For example, suppose that $\setP^{(n)}$ has only one AP pair where the low-power-receiving AP $a_f$ is on the left and high-power-receiving AP $a_c$ is on the right.
Here, the bilateration loss term would be equal to zero not only for an input position that is between $a_c$ and the midpoint of the two APs but also for another position that is to the right of $a_c$, leading to an ambiguity in accurately positioning the UE.
The loss function we propose next aims to mitigate such ambiguities as it also applies to UE positions with only one LoS AP.}

\subsection{LoS Bounding-Box Loss}
\label{sec:loss_box}

Assume that we have a bounding box as defined in~\fref{eq:boundingboxset} for the LoS area of each AP. The LoS bounding-box loss aims to place the UE inside the LoS bounding box of each AP which is estimated to have an LoS path to the UE.
Here, our CSI-based clue to the real-world position of the UE is the receive power at each AP:
If the receive power at an AP is sufficiently high, then one would expect the UE to be inside the LoS bounding box of this AP; this leads to the following \textit{LoS bounding-box loss}: 
\begin{align}
\boxloss= \, & \frac{1}{\sum_{n=1}^N|\tilde\setA^{(n)}|}  \sum_{n\in\setN} \sum_{a\in \tilde\setA^{(n)}}  \ell_{a}(g_{\boldsymbol\theta}(\vecf^{(n)})).\label{eq:loss_box}
\end{align}
Here,  we define $\ell_{a}(\vecx): \opR^{D} \to \opR$ {using the $a$th AP's rectangular LoS bounding box; e.g., for $D=2$ and $\vecx = [x \,\, y]^T$, we have}\footnote{The loss function written here has a simple form as we assume a rectangular LoS bounding box. This loss function can be modified to measure the distance from an arbitrary bounding box shape; investigating the effectiveness of such general bounding-box shapes is left for future work.\label{footnote_nonrectangular}}
\begin{align}
    \ell_{a}(\vecx) = \, & \mathds{1}_{\{x \not\in [ \xmin^{(a)},  \xmax^{(a)} ] \}}  \min_{x' \in \{\xmin^{(a)}, \xmax^{(a)} \}} |x'-x|^2  \notag \\ 
    & + \mathds{1}_{\{y \not\in [ \ymin^{(a)},  \ymax^{(a)}] \}} \min_{y' \in \{ \ymin^{(a)}, \ymax^{(a)} \}} |y'-y|^2 . \label{eq:loss_1a}
\end{align}
Intuitively, $\ell_{a}(\vecx)$ penalizes positions that are outside of the LoS bounding box~$\setB^{(a)}$: 
{the loss term should be non-zero only for coordinates that are \textit{not} in the given sets. In other words,} if~$\vecx$ is in the LoS bounding box~$\setB^{(a)}$, then $\ell_{a}(\vecx)$ is equal to zero; 
otherwise, the loss $\ell_{a}(\vecx)$ is equal to the minimum squared-distance to $\setB^{(a)}$. 
One can now train a neural network that implements the channel charting
function {$g_{\boldsymbol\theta}$} by minimizing the loss $\boxloss$; this approach is weakly supervised as the loss incorporates partial information on
the ground-truth labels (i.e., each AP's LoS bounding box).

We conclude by noting that a loss based on bounding boxes of zones was also proposed in~\cite{karmanov21wicluster}; however, this approach required zone-labeled CSI samples. 
In contrast, we deduce our bounding boxes solely based on the AP-side receive powers, without requiring any explicit CSI labels.

\subsection{Combining all Loss Functions}
\label{sec:loss_combined}

We propose {to use a linear combination of the three loss functions\footnote{{
We also experimented with applying the losses individually at different epochs, but this approach failed as the resulting channel chart seemed to be influenced primarily by the most recently applied loss function, as if the losses that were applied earlier had no impact at all. 
Therefore, we decided to apply all losses simultaneously in the form of a weighted sum in all epochs, which is straightforward with modern machine learning frameworks.}} \label{footnote_whysumloss}} $\tripletloss, \biloss$ and~$\boxloss$, since $\tripletloss$ helps to preserve local neighborhood relations whereas $\biloss$ and $\boxloss$ help to anchor the channel chart in real-world coordinates; 
moreover, false triplets in $\tripletloss$ and false AP pairs in $\biloss$ can be counterbalanced. 
We define the resulting multi-loss as follows: 
\begin{align} \label{eq:loss_combined}
\combinedloss = \lambdat \tripletloss + \lambdabi\biloss + \lambdabox\boxloss.  
\end{align}
Here, $\lambdat$, $\lambdabi$, and $\lambdabox$ are nonnegative regularization parameters that individually weight the losses; the choice of these parameters is briefly discussed in~\fref{sec:nn_structure}. 

One can train a neural network that implements the channel charting function $g_{\boldsymbol\theta}$ by minimizing the loss $\combinedloss$; this approach is a hybrid between self- and weakly-supervised learning due to the reasons mentioned in \fref{sec:triplet_cc},  \fref{sec:loss_bi}, and \fref{sec:loss_box}.



\section{Methods, Baselines, and Performance Metrics}
\label{sec:eval_setup}

We now briefly outline the neural network architecture used for learning the channel charting function. We then detail the proposed methods and baselines, and we introduce the performance metrics used in our evaluation.

\subsection{Learning the Channel Charting Function}
\label{sec:nn_structure}

We use a six-layer fully connected neural network, i.e., a simple multilayer perception (MLP), for the channel charting (and positioning) functions $g_{\boldsymbol\theta}$. 
With input feature dimension $D'$, the neural network has $\{D', D'/2, D'/4, D'/8, D'/16, D\}$ activations per layer,
and we set the output dimension to $D=2$ since the height of the UE is fixed.
All layers except for the last one use ReLU activations, whereas the last layer uses a linear activation function.
We employ Glorot initialization~\cite{glorot} for the last layer, and He initialization~\cite{kaiminginit} for the remaining layers.
We use the Adam optimizer during training. 

The hyperparameters used in the loss functions for training, namely~$\Tc$ and~$\Mt$ in the triplet loss from~\fref{eq:loss_t}, $\Mb$ in the bilateration loss from \fref{eq:loss_bi}, and the regularization parameters in the combined loss ($\lambdat$, $\lambdabi$, and $\lambdabox$) from~\fref{eq:loss_combined} affect the global scale of the resulting channel charts.
Unless stated otherwise, we keep $\lambdat=1$ and adjust the remaining hyperparameters.
We assume that the training dataset includes CSI from the entire area of interest in which the UEs are moving, and we choose these hyperparameters heuristically so that the resulting channel charts approximately match the size and shape of the area of interest.
{An example of how the hyperparameters interact is that if $\lambdabi$ and $\lambdabox$ are too small compared to $\lambdat$, then the channel chart lies in a relatively small absolute scale, far from real-world coordinates. 
Another example is that the smaller the margins $M_t$ and $M_b$ are, the smaller the absolute scale of the chart is (and vice versa). Given the clear correlation between hyperparameters and channel-chart size, adjusting the hyperparameters is rather  straightforward.}

\subsection{Proposed Methods and Baselines}
\label{sec:methods}

All of our proposed methods and baselines use the same neural network architecture (cf.~\fref{sec:nn_structure}) but are trained with different loss functions, which we detail next.

\subsubsection*{Proposed\,\,1 (P1)}
\label{sec:proposed1}

For our first proposed method, we extract the receive power for each UE position and each AP as described in \fref{sec:ccinreal}.
We train the neural network using the summation\footnote{\label{note1}For LoS bounding-box loss, we only used the LoS bounding box of the AP with the largest power since
we did not observe any improvement from using all APs in $\tilde\setA$ in our experiments.} of $\biloss$ and $\boxloss$ defined in~\fref{eq:loss_bi} and~\fref{eq:loss_box}, respectively; 
in other words, we use $\combinedloss$ defined in~\fref{eq:loss_combined} with $\lambdat=0$ and $\lambdabi,\lambdabox>0$.
Here, we aim to demonstrate the efficacy of only the bilateration and LoS bounding-box losses.
We reiterate that, due to the reasons mentioned in~\fref{sec:loss_bi} and~\fref{sec:loss_box}, this approach is weakly-supervised.

\subsubsection*{Proposed\,\,2 (P2)} 
\label{sec:combined_triplet_bilateration_loss}

For our second proposed method, we assume that timestamps are available for all CSI features in the training set; we also extract the receive power for each UE position at each AP as described in \fref{sec:ccinreal}.
We train the neural network using the multi-loss  $\combinedloss$ combining triplet, bilateration, and LoS bounding-box losses defined in~\fref{eq:loss_combined} with  $\lambdat,\lambdabi,\lambdabox>0$.
We reiterate that due to the reasons mentioned in \fref{sec:loss_combined} regarding each component of the loss, this approach is a hybrid between self- and weakly-supervised learning.

{
We note that, since we study more challenging scenarios than in the conference version of this paper~\cite{taner2023globecom},
we omit the proposed method from \cite{taner2023globecom} which uses the triplet and bilateration losses \textit{without} incorporating the LoS bounding-box loss.\footnote{
{For all scenarios that we study in this paper, our experimental results showed that the method from \cite{taner2023globecom} performs significantly worse than the P2 method {proposed here}; these results are in alignment with the limitations of {the} bilateration loss discussed in \fref{sec:loss_bi}.} \label{footnote_nooldmethod}}
} 

\subsubsection*{Baseline\,\,1 (B1)} 
\label{sec:onlytriplet}

For this baseline, we assume that
 the timestamps are available for all CSI features in the training set.
We train the neural network using only the triplet loss~$\tripletloss$ defined in \fref{eq:loss_t}.
Our aim in presenting this baseline is to compare the performance of our methods to (purely) self-supervised channel charting. 
We reiterate that this approach is self-supervised due to the reasons mentioned in \fref{sec:triplet_cc}.
Note that the resulting channel charts are often of high quality but embedded in arbitrary coordinates and are therefore unitless. This is in contrast to 
the weakly-supervised, semisupervised, and supervised methods
we present, which produce channel charts in real-world coordinates measured in meters.

\subsubsection*{Baseline\,\,2 (B2)} 
\label{sec:onlytriplet_affine}

For this baseline, we assume that, 
while timestamps are available for all CSI features in the training set,
ground-truth UE positions are available for a subset $\setS$ of the training set.
We first learn a channel charting function {$g_{\boldsymbol{\theta}}$} in a self-supervised manner in the same way as baseline B1; 
then, we estimate an affine transform that maps the output of~{$g_{\boldsymbol{\theta}}$} to real-world coordinates as in~\cite{pihljasalo20,stahlke23,stephan2023adp}  
\begin{align}
    \{\hat\bA, \hat\vecb\} = \argmin_{\bA\in\opR^{D\times D}, \bmb\in\opR^{D}} \,\, \sum_{n\in\setS}  \|(\bA g_{\boldsymbol\theta}(\vecf^{(n)}) + \vecb) - \tilde\vecx^{(n)}\|^2,
\end{align}
where we solve a least-squares problem to minimize the error between the affine-transformed channel chart positions and known ground-truth positions.
Finally, we apply the estimated affine transform to the channel-chart pseudo-positions as $\hat\vecx^{(n)} = \hat\bA g_{\boldsymbol\theta}(\vecf^{(n)}) + \hat\vecb$.
Our aim in presenting this baseline is to (i) showcase 
whether the channel charts can always be mapped to real-world coordinates through merely an affine transform, and 
(ii) compare the performance of our proposed method of incorporating the real-world coordinates directly into the channel charting function (without using ground-truth positions) to the two-stage approach of mapping a channel chart to real-world coordinates (using ground-truth positions).
We reiterate that estimating the affine transform here relies on known ground-truth positions.

\subsubsection*{Baseline\,\,3 (B3)} 
\label{sec:combined_triplet_mse_loss}

For this baseline, we assume
that, while the timestamps are available for all samples in the training set, ground-truth labels are available for a small subset $\setS$ of the training set.
Hence, we can define a mean-squared error (MSE) loss for the labeled samples as follows:
\begin{align}\label{eq:loss_mse}
    \loss_{\widetilde{\MSE}} = \frac{1}{|\setS|} \sum_{n\in\setS}  \|g_{\boldsymbol\theta}(\vecf^{(n)}) - \tilde\vecx^{(n)}\|^2.
\end{align}
We train the neural network using the combined triplet and MSE loss function defined as
\begin{align}
    \loss_{\mathrm{t},\widetilde{\MSE}}  = \loss_{\mathrm{t}}+\loss_{\widetilde{\MSE}} . \label{eq:loss_tMSE}
\end{align}
Since the loss $\loss_{\mathrm{t},\widetilde{\MSE}}$ utilizes a subset of the ground-truth UE positions in addition to the triplet loss, this baseline is semi-supervised.
Our aim in presenting this baseline is to compare the performance of our methods to semi-supervised positioning techniques (e.g., \cite{lei19siamese,penghzi19spawc,deng21networkside,karmanov21wicluster,zhang21globecom,deng21tsne}), 
and showcase whether our approach can achieve comparable performance to semi-supervised techniques without using any ground-truth UE position information. 

\subsubsection*{Baseline\,\,4 (B4)} 
\label{sec:sup}

For this baseline, we assume that {ground-truth positions} are available for all samples in the training set.
We train the neural network with the MSE loss for all training samples.
We denote the loss here by $\loss_{\MSE}$ to distinguish it from the loss over a subset {of} training samples as in baseline B3.
Since the loss $\loss_{\MSE}$ requires ground-truth labels for \emph{all} training samples, this approach is fully supervised. 
Our aim in presenting this baseline is to compare the performance of our methods to fingerprinting methods that process CSI features with neural networks (see, e.g.,~\cite{ferrand2020globecom,gonultas22twc} 
and the references therein).
This baseline is also an indicator {of} the best-possible positioning performance achievable with the neural network architecture described in~\fref{sec:nn_structure}.

\subsection{Performance Metrics}
\label{sec:perf_metrics}
We evaluate the performance of the proposed methods and baselines using six metrics.
The first four metrics measure the quality of the channel chart.
The latter two metrics measure positioning accuracy.
In the following, we denote the set of the $J$ closest neighbors of each real world position $\tilde\bmx^{(n)}$ and each latent space position $\hat\bmx^{(n)}$ by $\tilde\setV_n $ and $\hat\setV_n $, respectively.
We denote the ranking of how close $\tilde\bmx^{(j)}$ is to $\tilde\bmx^{(n)}$ in the real world and $\hat\bmx^{(j)}$ is to $\hat\bmx^{(n)}$ in the latent space in terms of Euclidean distance by  $\tilde{r}(n,j)$ and  $\hat r(n,j)$, respectively.
We define $\gamma = \frac{2}{NJ(2N-3J-1)}$ as a normalization factor \cite{altous22asilomar}.

\subsubsection*{Trustworthiness (TW)} 
This {metric} penalizes neighborhood relationships in latent space that are not present in the real-world positions {and is defined as~\cite{vathy2013}
\begin{align} \label{eq:TW}
\textit{TW}(J) = 1-\gamma \sum_{n=1}^N \sum_{\substack{j \notin \tilde \setV_n \\ j \in \hat \setV_n}} ({\tilde{r}(n,j)}-J).
\end{align}
The TW is in $[0,1]$ with optimal value~$1$}. 

\subsubsection*{Continuity (CT)} This  {metric} measures how well neighborhood relationships between the real-world positions are preserved in latent space
{and is defined as~\cite{vathy2013}
\begin{align} \label{eq:CT}
\textit{CT}(J) = 1-\gamma \sum_{n=1}^N \sum_{\substack{j \in \tilde \setV_n \\ j \notin \hat \setV_n}} ({\hat{r}(n,j)}-J).
\end{align}
The CT is in $[0,1]$ with optimal value~$1$}. 

\subsubsection*{Kruskal stress (KS)} This {metric} measures the dissimilarity between pairwise distances {$\tilde d(n,j)\define \|\tilde\bmx^{(n)}-\tilde\bmx^{(j)}\|$} in the real-world positions {$\big\{\tilde\bmx^{(n)}\}_{n=1}^N$} and pairwise distances {$\hat d(n,j)\define\|\hat\bmx^{(n)}-\hat\bmx^{(j)}\|$} in latent space {positions} {$\big\{\hat\bmx^{(n)}\}_{n=1}^N$}{ and is defined as~\cite{kruskal1964multidimensional}
\begin{align} \label{eq:KS}
\textit{KS} = \min_{\lambda\in\reals} \sqrt{\frac{\sum_{n=1,j=1}^N \big(\hat d(n,j)-\lambda \tilde d(n,j)\big)^2}{\sum_{n=1,j=1}^N \hat d(n,j)^2}},
\end{align}
where $\lambda$ is a scaling factor. The KS is in $[0,1]$ with optimal value~$0$.} 

\subsubsection*{Rajski distance (RD)} This {metric} measures the difference between the mutual information and joint entropy of the distribution of pairwise distances in the real-world positions and latent space, {$\tilde d(n,j)$ and $\hat d(n,j)$, respectively, and is defined as~\cite{rajski1961}
\begin{align} \label{eq:RD}
\textit{RD} = 1-\frac{I(V,Q)}{H(V,Q)} \text{, for } H(Q,V) \neq 0,
\end{align}
where $V$ and $Q$ are two discrete random variables with mutual information given by
\begin{align} \label{eq:mutual_information}
    I(V,Q) = \sum_{v \in V, q \in Q} P_{V,Q}(v,q) \log_2 \frac{P_{V,Q}(v,q)}{P_V(v)P_Q(q)}
\end{align}
and joint entropy information given by
\begin{align} \label{eq:joint_entropy}
    H(V,Q) = - \sum_{v \in V, q \in Q} P_{V,Q}(v,q) \log_2 P_{V,Q} (v,q). 
\end{align}
In \fref{eq:mutual_information} and \fref{eq:joint_entropy}, $P_{V,Q}(v,q)$ is the joint probability distribution, and $P_V(v)$ and $P_Q(q)$ are the marginal distributions of the quantized pairwise distances in the original $\tilde d(n,j)$ and latent space $\hat d(n,j)$. 
The RD is in $[0,1]$ with optimal value~$0$.} 

\subsubsection*{Mean distance error (MDE)} This {metric} measures the average error in the UE position estimates in  Euclidean norm over all UE positions {and is defined as
\begin{align}
\textit{MDE} = \frac{1}{N}\sum_{n=1}^N\|\hat\bmx^{(n)}-\tilde \bmx^{(n)}\|.
\end{align}
The MDE} is nonnegative with optimal value~$0$.

\subsubsection*{95th percentile distance error} 

{This metric represents the 95th percentile of the Euclidean-norm errors in UE position estimates, defined as the smallest value $e_{95}$ such that at least 95\% of the distance errors $ \{\|\hat\bmx^{(n)}-\tilde \bmx^{(n)}\|\}_{n\in\setN}$ are less than~$e_{95}$.}



\section{Results}
\label{sec:results}

We assess the performance of our proposed methods and the considered baselines using three distinct CSI datasets obtained from different scenarios: a simulated outdoor scenario, a simulated indoor scenario, and a measurement-based indoor scenario.
In the subsections below, we describe each scenario,
provide a brief analysis of the AP-side receive power,
and show the ground-truth position, channel chart, and estimated position plots, followed by assessing the performance metrics from~\fref{sec:perf_metrics} for the proposed methods and baselines.

In the simulated scenarios, we add noise to the truncated channel matrices so that the expected value of the highest signal-to-noise ratio (SNR) per AP is $25$\,dB {with the goal of modeling channel estimation errors that arise in realistic communication systems. Concretely, we determine the additive noise variance so that 
the average SNR for the truncated channel matrix of each AP with the highest Frobenius norm
is $25$\,dB. This approach results in much lower SNR values for most UE positions.}
In all three scenarios, we use $80\%$ of the dataset for training and $20\%$ for testing.
For baselines B2 and B3, we select the subset of ground-truth labeled CSI samples uniformly at random from the training set.
All plots and performance metrics are based on the test set.

In Figs.~\ref{fig:outdoor_results}, \ref{fig:indoor_results}, and~\ref{fig:dichasus_results}, we visually compare the ground-truth
positions as well as channel charts in arbitrary and real-world
coordinates for each scenario;
in Tbls.~\ref{tbl:outdoor}, \ref{tbl:indoor}, and~\ref{tbl:dichasus}, we provide the respective performance metrics.
The performance metrics reflect the average based on 10 random initializations of the neural network parameters.
{When evaluating the TW and CT metrics, we set $J=0.05N$ as in~\cite{studer18cc}; when evaluating the RD metric, we quantize the pairwise distances into $20$ uniform bins.}
For the positioning metrics, we also report the standard deviation; we omitted this step for the latent space quality metrics since these metrics are all between $[0,1]$ which results in small standard deviations.

Since B1 produces channel charts in arbitrary coordinates, we omit the two distance error metrics for baseline B1 in the performance metric tables. Moreover, to prevent redundancy, we exclude the figures depicting estimated positions for supervised learning baseline B4, as these figures closely resemble those of the ground-truth positions.

\begin{table*}
\centering
\caption{Parameters for each Considered Scenario}\label{tbl:simulation_parameters}
\begin{tabular}{@{}llll@{}}
\toprule
 & \multicolumn{3}{c}{Value or type of parameter for each scenario} \\
\cmidrule(){2-4}
Parameter & Simulated outdoor & Simulated indoor & Measured indoor~\cite{dataset-dichasus-cf0x} \\
\midrule
Number of APs & $A=6$  & $A=8$ & $A=4$ \\
Number of antennas per AP & $M_R=4$ & $M_R=4$  & $M_R=8$  \\
APs' antenna array structure & Uniform linear array & Uniform linear array  & Uniform rectangular array, $2\times 4$  \\
Number of UE positions & $14\,642$ & $14\,606$ & $16\,778$ \\
Spacing between UE positions &40\,cm & 2.5\,cm & (not fixed) \\
AP antenna height & $10$\,m & $2$\,m  & $1.56-1.78$\,m \\
AP antenna spacing & Half-wavelength &  Half-wavelength  &  Half-wavelength  \\
UE antenna height & $1.5$\,\text{m} & $1.5$\,\text{m} & $\approx 0.94\,\text{m}$ \\
AP and UE antenna type & Omnidirectional &  Omnidirectional & Omnidirectional \\
Carrier frequency & $1.9$\,GHz & $2.4$\,GHz & $1.272$\,GHz \\
Bandwidth & $20$\,MHz & $20$\,MHz & $50$\,MHz \\
Number of used subcarriers & $W=1200$ & $W=52$ & $W=1024$\\
Maximum SNR per AP & $25$\,dB  & $25$\,dB & -- \\
\bottomrule
\end{tabular}
\end{table*}

\subsection{Simulated Outdoor Scenario}

\subsubsection{Description}
\label{sec:outdoor_description}

For our first scenario, we consider a \mbox{D-MIMO} urban environment
in which {six} APs 
are distributed around a rectangular area of size $83\,\text{m} \times 122\,\text{m}$, while the UE traverses a trajectory to cover this region with north-south and east-west meandering; 
see {\fref{fig:scenario_outdoor}} for a snapshot of the simulation environment, and {\fref{fig:outdoor_gt}} for an illustration of the UE and AP positions along with the LoS bounding box used for all APs.
We simulate a sub-6-GHz scenario with channel vectors from Remcom's Wireless InSite ray-tracing software \cite{remcom}.
The first column of \fref{tbl:simulation_parameters} summarizes the simulation parameters.\footnote{This simulation scenario is slightly different from what we presented in our conference paper~\cite{taner2023globecom}: 
In~\cite{taner2023globecom}, we used a rectangular grid for the UE positions and eight APs; 
here, we used the trajectory feature of Remcom's Wireless In-Site for the UE positions and six APs for a more challenging scenario.}
Each CSI matrix $\bH^{(n)}$, $n\in\setN$ is of dimension $24\times1200$; each CSI feature vector is of dimension $D'=24\times8=192$, where we use the $C=8$ first taps{\footnote{{In our experiments, we observed that further increasing the number of used channel taps led to only a slight improvement in the considered performance metrics, while significantly increasing the number of parameters of the neural network and the resulting computational complexity.} \label{footnote_numtaps}}} and the feature extraction pipeline outlined in~\fref{sec:feature_extraction}.

\begin{figure}[tp]
    \centering
    \subfigure[Outdoor scenario]
    {
    \includegraphics[height=3.1cm]{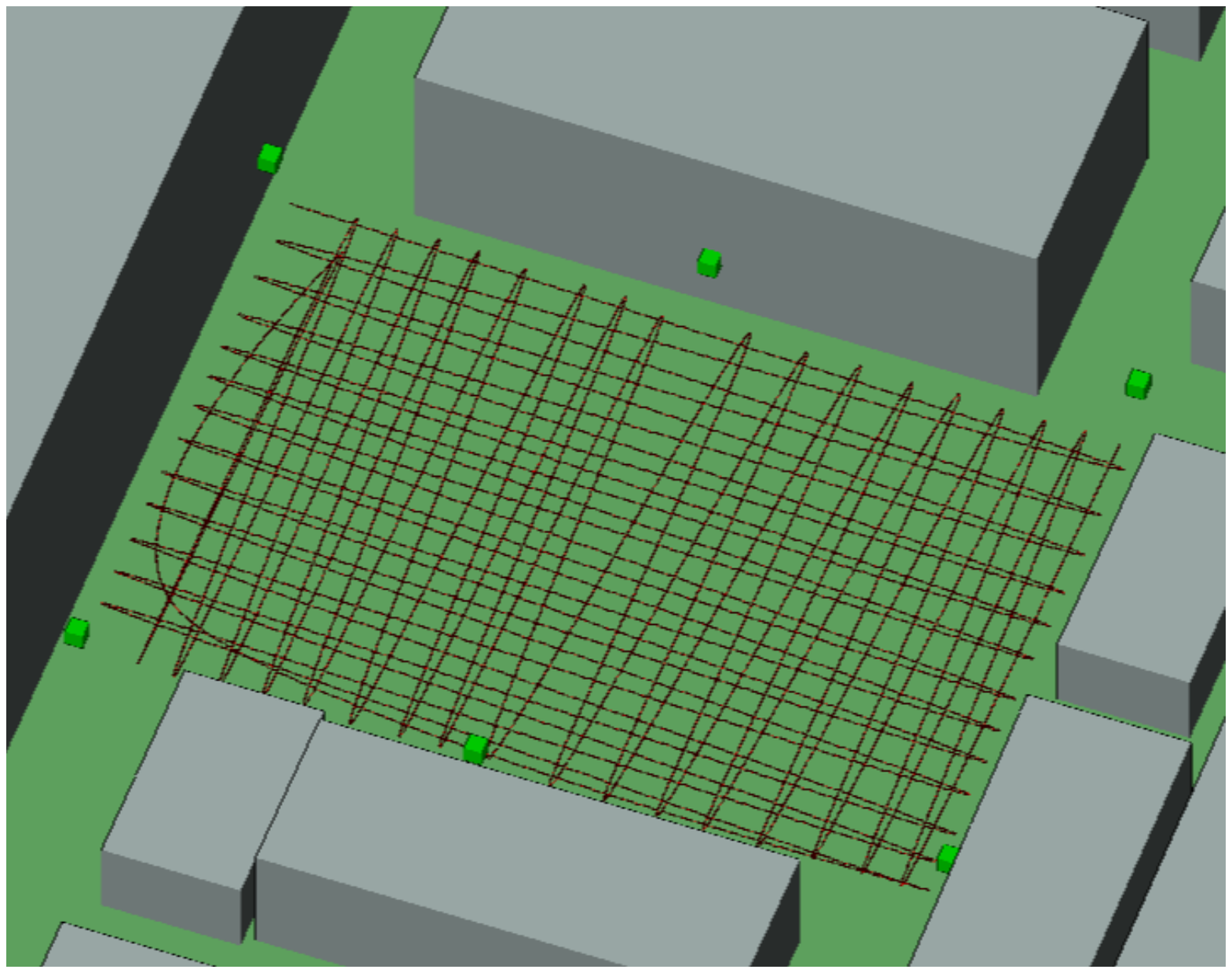}
    \label{fig:scenario_outdoor}
    }
    \subfigure[Indoor scenario]
    {
    \includegraphics[height=3.1cm]{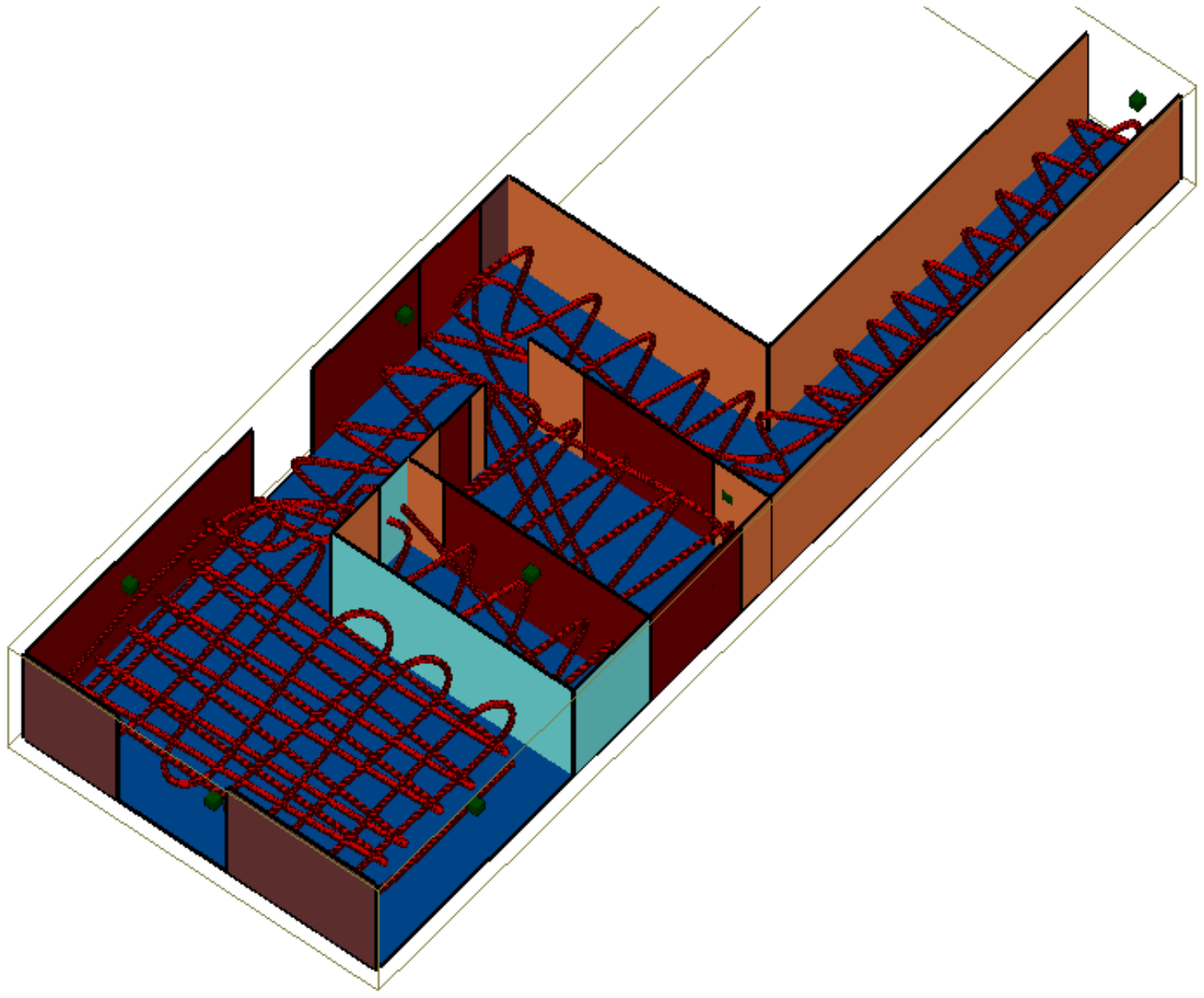}
    \label{fig:scenario_indoor}
    }
    \caption{Simulated D-MIMO scenarios: (a) outdoor scenario with $14\,642$ UE positions and $A = 6$ APs; (b) indoor scenario with $14\,606$ UE positions and $A = 8$ APs. Each AP (green boxes) is equipped with $M_R = 4$ antennas.} \label{fig:scenarios_snapshots}
\end{figure}

\subsubsection{Analysis of AP-Side Receive Power}
\label{sec:outdoor_power}

In this scenario, we know that all APs have a LoS path to all UE positions; therefore, we can set $\Pthr=-\infty$ to include all APs in $\tilde\setA^{(n)}$ from \fref{eq:tildeA}.
We exemplify the power-distance relation exploited by bilateration loss in \fref{fig:outdoor_pow_distance}.
Here, we show the receive power at one AP depending on the distance to the UE in the xy-plane; we ignore the distance in the z-axis since the heights are fixed. 
{We remark that our proposed methods do not use ground-truth positions, and thus, are uninformed of the UE's distance to APs; we do not make any decisions based on the power-distance plot and we merely provide this plot as a proof-of-concept.}
Evidently, the power generally decreases with distance, but small-scale fading causes the power to fluctuate quite substantially among the different UE positions which are at approximately the same distance to the AP.
Furthermore, the received power increases\footnote{This behavior can be attributed to the antenna radiation pattern, which causes the receive power to reduce if a UE approaches the area below an AP.} with distance until about $10$\,m, which negatively affects approaches that rely on the power being the highest near the AP, e.g., the method in~\cite{pihljasalo20}. 

Since a smaller difference in power is more likely to be caused by small-scale fading (and not by distance), we would be less confident in deducing which AP might be closer. Hence, 
observing that the AP-side receive power is not perfectly inversely proportional to distance motivates the margin parameter $\Mp$ in \fref{eq:setP}.
By setting $\Mp>0$, we can avoid some false AP pairs, i.e., pairs $(a_c,a_f) \in \setP^{(n)}$ for which $\|\tilde\vecx^{(n)}- \tildexap{a_c}\| > \|\tilde\vecx^{(n)}- \tildexap{a_f}\| $.
For this CSI dataset, margins $\Mp$ of $0$, $3$, and $6$ result in false AP pair ratios of  $16.6$\%, $0.06$\%, and $0.02$\% for all UEs, 
while the average number of AP pairs in the set $\setP^{(n)}$ given by $\frac{1}{N}\sum_{n=1}^N |\setP^{(n)}|$ is $15$, $9.2$, and $4.63$, respectively.
Clearly, there exists a trade-off in choosing~$\Mp$:
Increasing $\Mp$ leads to fewer \emph{false} AP pairs at the cost of fewer AP pairs in $\setP^{(n)}$ from \fref{eq:setP},
which yields more accurate but fewer reference points for weakly-supervised learning (and vice versa).

\begin{figure}[tp]
\centering
\includegraphics[width=0.9\columnwidth]{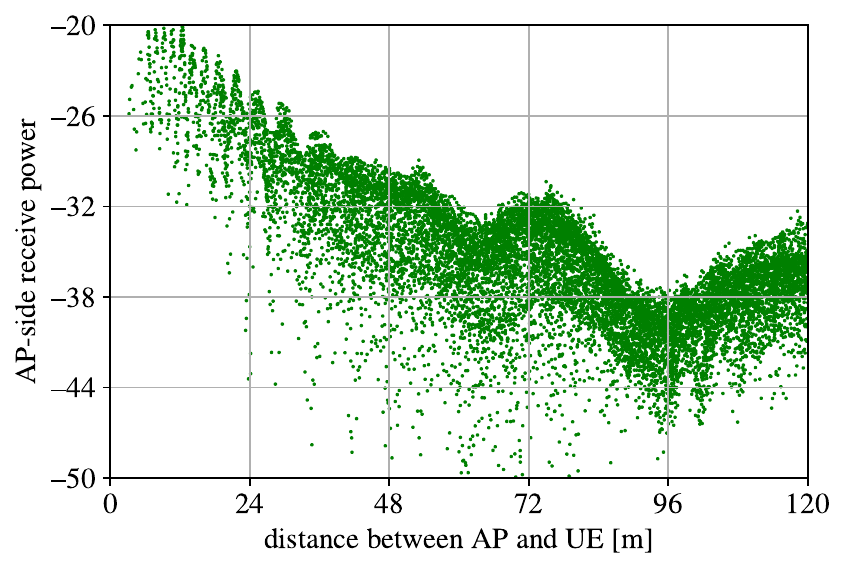}
\vspace{-0.15cm}
\caption{Receive power (in decibels) at one AP for the simulated outdoor scenario with respect to the AP-to-UE distance in the xy-plane.} 
\label{fig:outdoor_pow_distance}
\end{figure}

\subsubsection{Parameter Settings}
\label{sec:outdoor_param}

For our proposed method P1, we set $\Mt=\Mb=10$,  $\Mp=3$, $\lambdat = 0$, and $\lambdabi = \lambdabox=1$.
Since there is no blockage in the area of interest, i.e., the UEs should have a LoS path to all APs, we set $\Pthr=-\infty$.
P2 shares the same parameters as P1 except for the triplet loss, for which we use $\lambdat = 1$, $\Tc=10$, and $\Mt=10$.
B1, B2, and B3 share the same triplet loss parameters as P2.

For the baselines B2 and B3, we used the ground-truth positions of $10$ and $400$ CSI samples, respectively.
With B2, we demonstrate the performance of learning an affine transform from a handful of labeled samples.
With B3, we demonstrate approximately how many ground-truth labeled samples are required to achieve comparable performance to P2.

\subsubsection{Performance Evaluation}
\label{sec:outdoor_eval}

In \fref{fig:outdoor_results}, we observe 
that all methods result in an almost rectangular chart preserving the color gradient and resembling the ground-truth positions, as desired.   
Potentially due to the shape of the area and the trajectory of the UE, the self-supervised baseline B1 performs surprisingly well in terms of the global geometry; the channel chart only needs to be rotated, which is exactly what B2 yields.

In \fref{tbl:outdoor}, we observe that 
affine transform-based B2 has the best performance in all metrics, even outperforming the supervised baseline B4 {by $0.004$, $0.04$, $0.005$, and $0.030$ in TW, CT, KS, and RD, respectively}. 
The reason behind the performance of B2 is that the self-supervised baseline B1 luckily already represents both the local and global geometry very well, as confirmed by the large values of {$0.998$ for both} TW and CT, and relatively small values of {$0.068$ and $0.550$ for} KS and RD, respectively.
Hence, we note that \textit{when} the channel chart represents the global geometry well, then an affine transform learned from a handful of labeled samples can result in a good positioning performance.\footnote{Here, we have observed no  significant improvement from further increasing the number of labeled samples.}
However, the shortcomings of B2 will be apparent in our next scenario.
While the channel chart of P1 demonstrates that the bilateration and LoS bounding-box losses alone can help create a channel chart in real-world coordinates,
P1 is outperformed by all the other methods in all performance metrics.
P2 outperforms the semi-supervised baseline B3 in the latent space quality metrics TW, CT, KS, and RD by $0.020$, $0.023$, $0.043$, and $0.065$, respectively.
P2 is slightly better than B3, by approximately $0.3$\,m in mean and $2$\,m in the 95th percentile,\footnote{{In \cite{taner2023globecom}, our simulation scenario included eight APs in the same environment. 
Here, we make our simulation scenario more challenging by decreasing the number of APs to six, which requires the use of our novel LoS bounding-box loss in order to maintain the positioning performance of our methods.}\label{footnote_sixap}} while holding the advantage of not requiring any ground-truth positions. 
P2 has a higher positioning error than B2 by approximately $5$\,m in mean and $11$\,m in the 95th percentile.

We reiterate that the proposed method P2 does not require any ground-truth UE position information as opposed to baselines B2, B3, and B4.
It is remarkable that B3 would require \textit{at least} 400 ground-truth labeled samples to achieve the same performance as P2.
These results demonstrate the efficacy of our proposed method P2 in a large, outdoor area with LoS channels.

\newcommand{\figsize}{0.31}

\begin{figure*}[tp]
    \centering
    \subfigure[Ground-truth UE positions]
    {    \includegraphics[width=\figsize\textwidth]{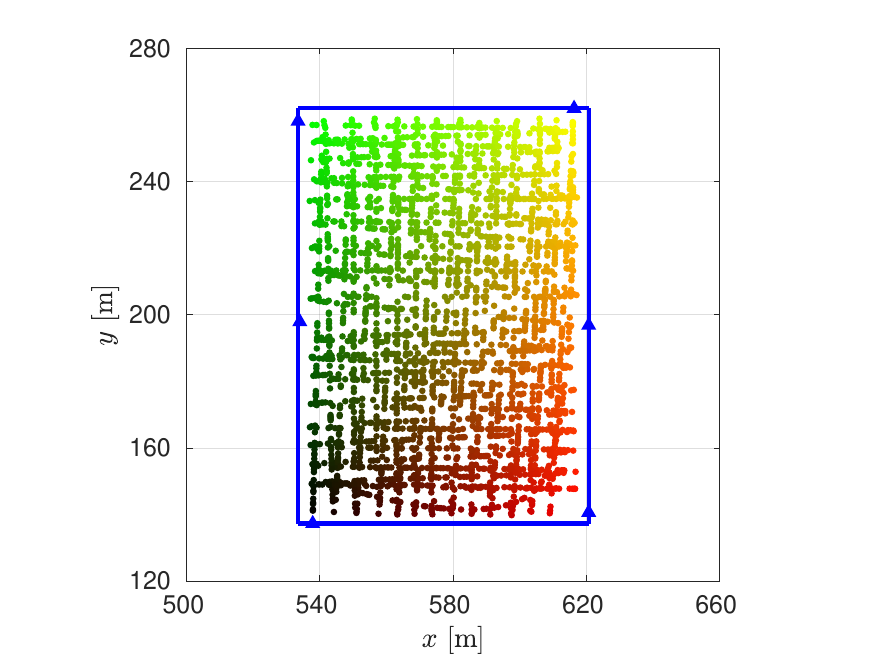}
    \label{fig:outdoor_gt}
    }
    \subfigure[P1 (weakly-supervised)]  
	{
    \includegraphics[width=\figsize\textwidth]{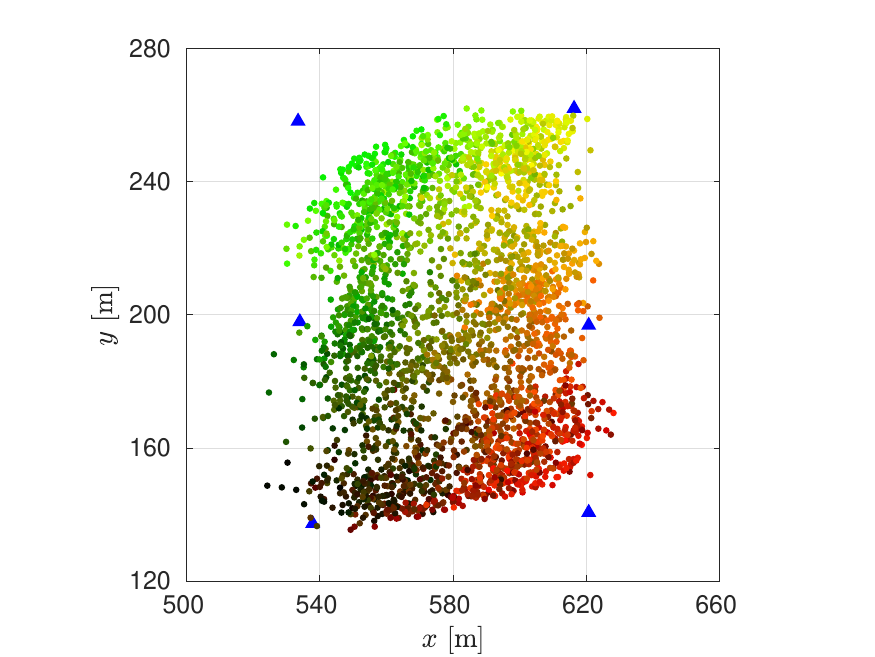}
    \label{fig:outdoor_p1}
    }
    \subfigure[P2 (hybrid self- and weakly-supervised)]
	{
    \includegraphics[width=\figsize\textwidth]{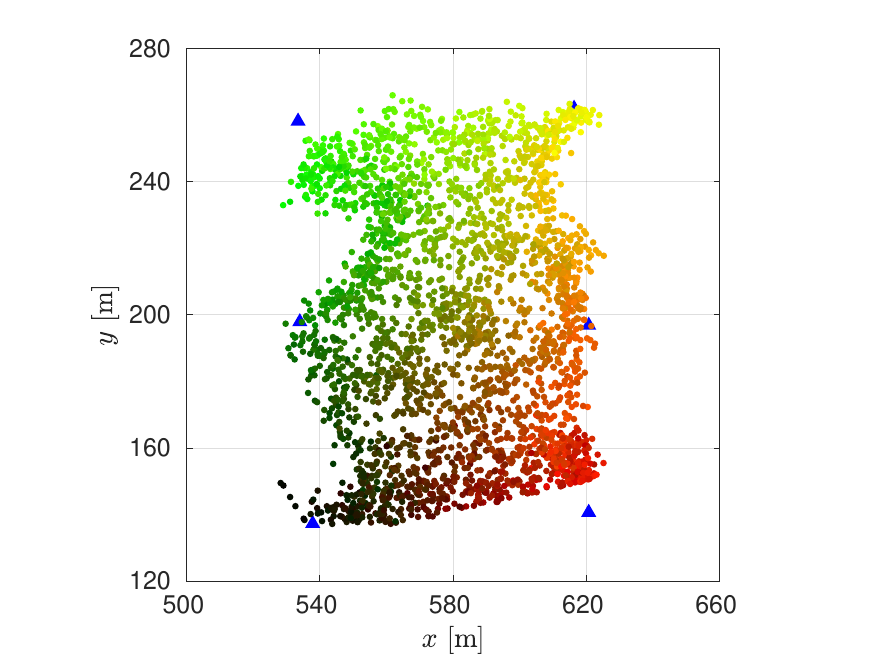}
    \label{fig:outdoor_p2}
    }
     \subfigure[B1 (self-supervised)]
	{
    \includegraphics[width=\figsize\textwidth]{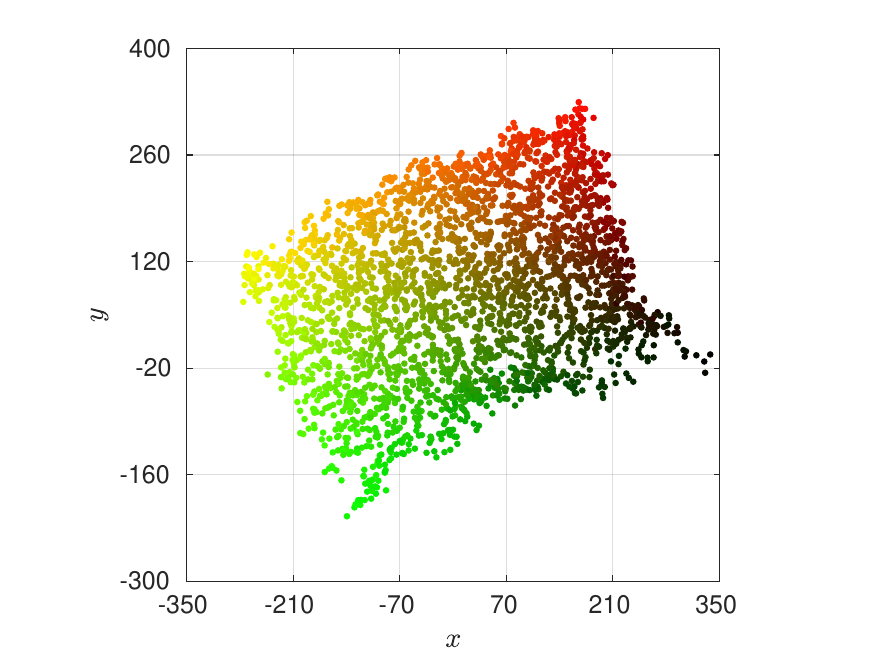}
    \label{fig:outdoor_triplet}
    }
    \subfigure[B2 (affine transformed B1)]
	{
    \includegraphics[width=\figsize\textwidth]{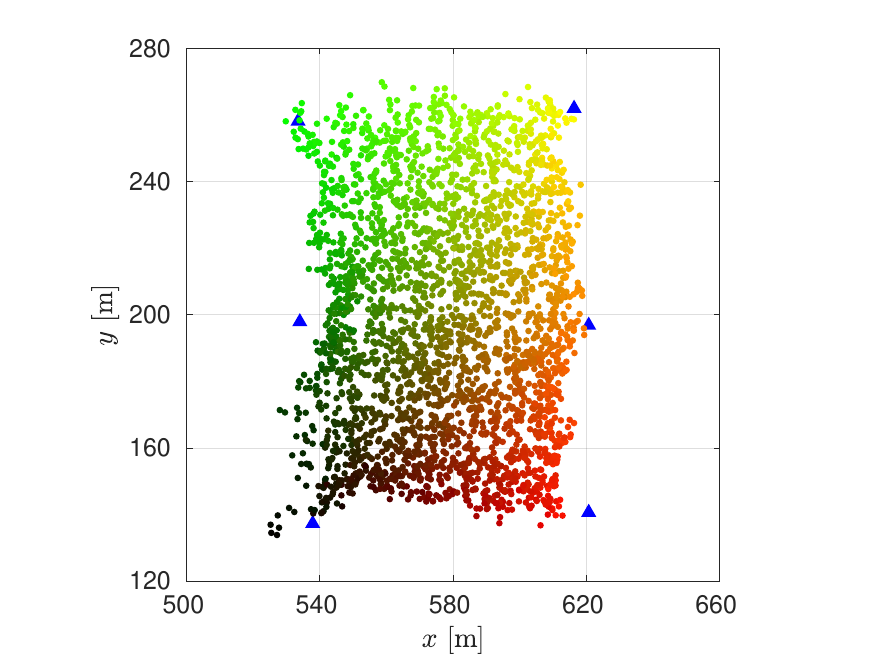}
    \label{fig:outdoor_affine}
    }
    \subfigure[B3  (semi-supervised)]
	{
    \includegraphics[width=\figsize\textwidth]{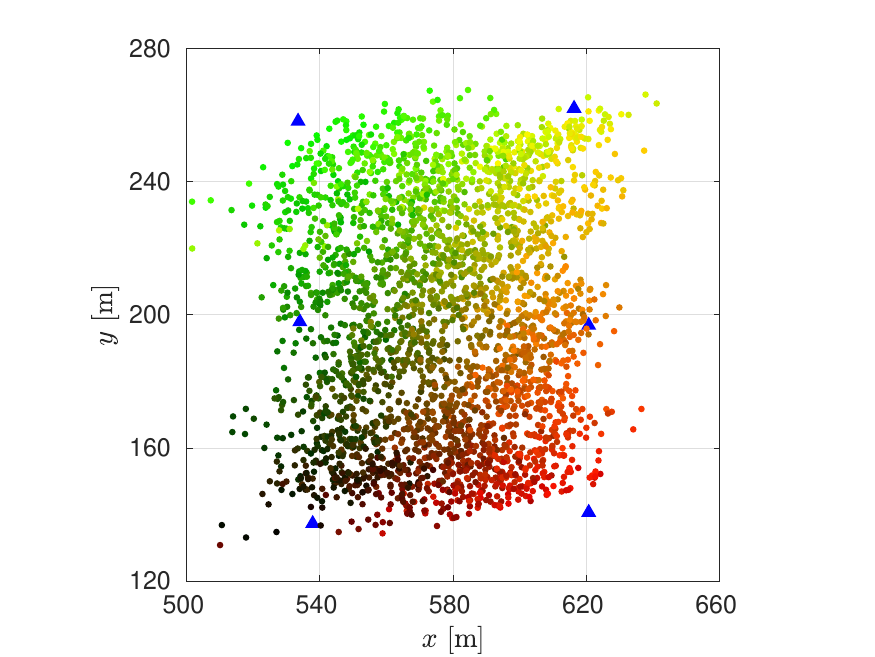}
    \label{fig:outdoor_semisup}
    }
    \caption{Results for the simulated outdoor scenario: (a) ground-truth UE positions (green-to-red gradient colored area), AP positions (blue triangles), and the LoS bounding box of all APs; and (b-f) the channel charts or positioning estimates for the proposed (P) and baseline (B) methods. Since the output of baseline B2 is in arbitrary coordinates, the AP positions are not shown in (d). The proposed method P2 achieves comparable results in real-world coordinates as the semi-supervised baseline B3 but without requiring known UE positions during training. } 
    \label{fig:outdoor_results}   
\end{figure*}

\begin{table*}
\centering
\resizebox{0.92\textwidth}{!}{
\begin{minipage}[c]{0.99 \textwidth}
    \centering
    \caption{Channel charting and positioning performance comparison for the simulated outdoor scenario}
    \label{tbl:outdoor}		 
    \begin{tabular}{@{}lccccccc@{}}
        \toprule
        && \multicolumn{4}{c}{Latent space quality metrics} & \multicolumn{2}{c}{Positioning error [m]} \\
        \cmidrule(lr){3-6} \cmidrule(){7-8}  
        Method  & Figure & TW$\,\uparrow$ & CT$\,\uparrow$ & KS$\,\downarrow$ & RD$\,\downarrow$  &  Mean$\,\downarrow$ & 95th percentile$\,\downarrow$ \\
        \midrule
        P1 & {\ref{fig:outdoor_p1}} & 0.923  & 0.922  & 0.238  & 0.864& 14.18 $\pm$ 0.62 & 29.98 $\pm$ 1.306  \\ 
        P2 & {\ref{fig:outdoor_p2}} & 0.984  & 0.985 & 0.123  & 0.710  & 8.75 $\pm$ 0.31 & 17.85 $\pm$ 0.66 \\ 
        \midrule
        B1 & {\ref{fig:outdoor_triplet}} & 0.998 & 0.998 & 0.068  & 0.550 & -- & -- \\
        B2 & {\ref{fig:outdoor_affine}} & 0.998  & 0.998 & 0.064 & 0.533 & 3.15 $\pm$ 0.34 & 7.06 $\pm$ 0.99 \\  
        B3 &  {\ref{fig:outdoor_semisup}}  & 0.964  & 0.962  & 0.166  & 0.775 & 9.07 $\pm$ 0.47 & 20.76 $\pm$ 1.28   \\ 
        B4 &  --  & 0.994  & 0.994  & 0.069  & 0.563 & 3.57 $\pm$ 0.18 & 7.34 $\pm$ 0.37   \\
        \bottomrule
    \end{tabular}		 
\end{minipage}}
\end{table*}


\subsection{Simulated Indoor Scenario}
\label{sec:indoor}

\subsubsection{Description}
\label{sec:indoor_description}

For our second scenario,  we consider a challenging indoor environment where many APs are in non-LoS for a given UE position {because of interior walls}.
At least one AP is in the LoS of the UE while the UE traverses a trajectory to cover each hallway and room; 
see  {\fref{fig:scenario_indoor}} for a snapshot of the simulation environment, and 
see {\fref{fig:indoor_gt}} for an illustration of the UE and AP positions along with three APs' LoS bounding boxes as examples;
here, we omit to show the LoS bounding boxes of the remaining five APs to avoid crowding the figure.
We simulate a sub-6-GHz scenario with channel vectors from Remcom's Wireless InSite ray-tracing software \cite{remcom}. 
The second column of~\fref{tbl:simulation_parameters} summarizes the simulation parameters.
Each CSI matrix $\bH^{(n)}$, $n\in\setN$, is of dimension $32\times52$; each CSI feature vector is of dimension $D'=32\times8=256$, where we use the $C=8$ first taps and the feature extraction pipeline from \fref{sec:feature_extraction}.

\subsubsection{Analysis of AP-Side Receive Power}
\label{sec:indoor_power}

In this scenario, we know that the UE does not have a LoS path to some APs at each position; therefore, we need to choose a power threshold to estimate the LoS APs in $\tilde\setA^{(n)}$ from \fref{eq:tildeA}.
We choose this threshold based on the AP-side receive power $P^{(n,a)}$ over time $t_n$, as exemplified for one of the APs in~{\fref{fig:indoor_pow_time}}.
Here, we observe a steep drop in the power at around $-30$\,dB, and set $\Pthr=-30$.{\footnote{{
For simplicity, we decided to set the power threshold based on a visual inspection of the received power over time. Alternatively, one could sort the receive power values in ascending order and apply an edge detection algorithm to identify the transition between the non-LoS and LoS regions. We leave an exploration of such automated parameter-selection approaches for future work.} \label{footnote_threshold}}} 
We omit an ablation study that shows $\Pthr$ is robust to small deviations for brevity.
We now exemplify the power-distance relation exploited by our bilateration loss in \fref{eq:loss_bi} and the impact of $\Pthr$ for the same AP as~{\fref{fig:indoor_pow_time}} in~{\fref{fig:indoor_pow_distance}}.
Here, we clearly observe the need for choosing an appropriate power threshold $\Pthr$ as many UE positions are close to the AP while the AP-side receive power is low;
we should not use this AP in bilateration loss for such UE positions. 
As desired, we observe that the receive power indeed falls below our chosen power threshold $\Pthr$ for these positions.
Moreover, the UE positions where the AP-side receive power is higher than $\Pthr$  follow the same trend as \fref{fig:outdoor_pow_distance}, i.e., the power generally decreases with distance.
Therefore, we conclude that our choice of $\Pthr$ is suitable for classifying the LoS APs to be used in bilateration and LoS bounding-box losses.

For this CSI dataset, a threshold $\Pthr=-30$ with margins  $\Mp$ of $0$, $3$, and $6$ result in false AP-pair ratios of  $15.5$\%, $3.4$\%, and $0.2$\% for all UEs, 
while the average number of AP pairs in the set $\setP^{(n)}$ given by $\frac{1}{N}\sum_{n=1}^N |\setP^{(n)}|$ is $1.03$, $0.60$, and $0.30$, respectively.
Once again, we observe the trade-off in choosing~$\Mp$: Increasing $\Mp$ leads to fewer false AP pairs at the cost of fewer AP pairs in $\setP^{(n)}$ from \fref{eq:setP}.

\begin{figure}
\centering
\subfigure[AP-side receive power with respect to time]
{
\includegraphics[width=0.9\columnwidth]{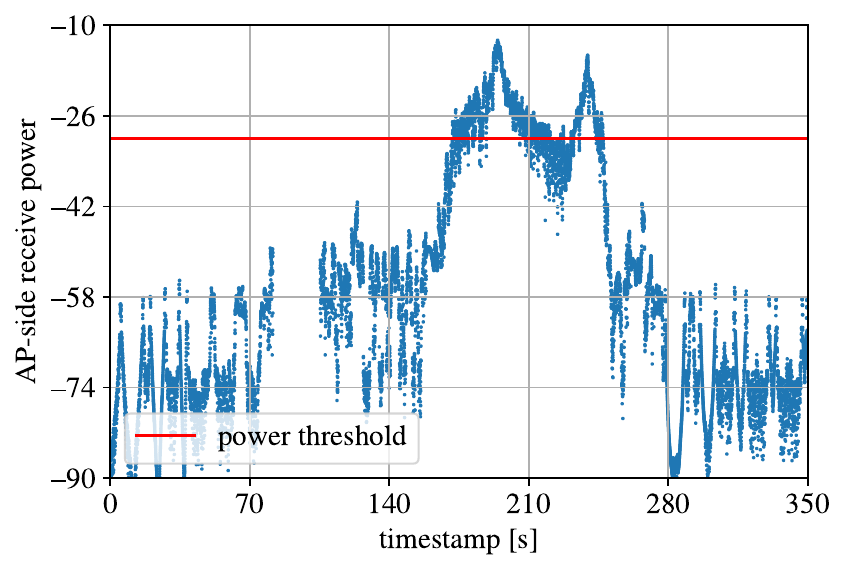}
\label{fig:indoor_pow_time}
}
\subfigure[AP-side receive power versus AP-to-UE distance]
{
\includegraphics[width=0.9\columnwidth]{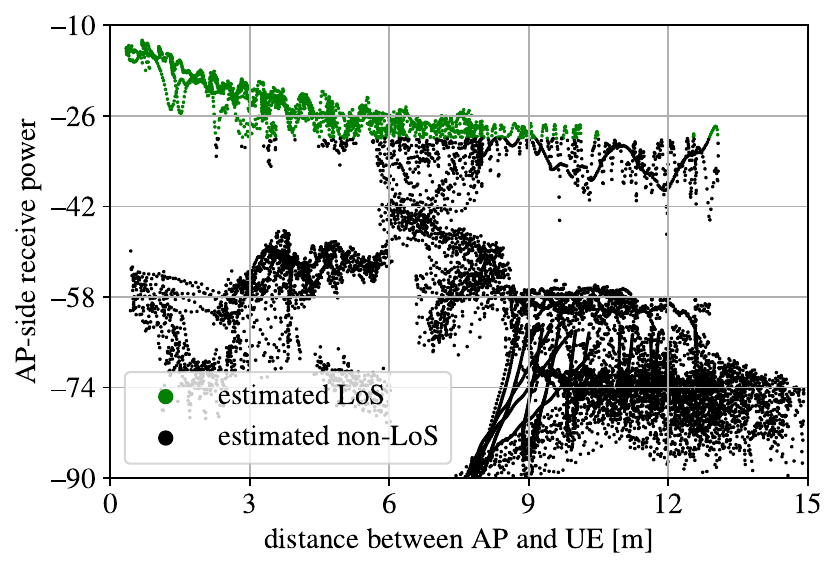}
\label{fig:indoor_pow_distance}
}
\caption{Receive power (in decibels) at one AP for the simulated indoor scenario with respect to (a) timestamps and (b) AP-to-UE distance in the xy-plane. In (a), the red line designates the power threshold chosen to identify LoS APs. In (b), the power values above the threshold (highlighted with green) indicate instances where the AP is estimated to be in LoS.} 
\label{fig:indoor_pow}
\end{figure}

\subsubsection{Parameter Settings}

For our proposed method P1, we set $\Mt=\Mb=1$,  $\Mp=3$, $\lambdat = 0$, $\lambdabi = \lambdabox=1$, and $\Pthr = -30$.
P2 shares the same parameters as P1 except for the triplet loss, for which we use $\lambdat = 1$, $\Tc=2$, and $\Mt=1$.
B1, B2, and B3 share the same triplet loss parameters as P2.

For the baselines B2 and B3, we used the ground-truth positions of $10\,000$ and $50$ CSI samples, respectively.
With B2, we demonstrate that the affine transform fails to yield a good representation of real-world coordinates no matter how many ground-truth positions were used.
With B3, we demonstrate approximately how many ground-truth labeled samples are required to achieve comparable performance to P2.

\subsubsection{Performance Evaluation}
\label{sec:indoor_eval}

In \fref{fig:indoor_results}, we observe that P1, P2, and B3 all yield similar results while the quality of P1 is (unsurprisingly) slightly worse. However, even in P1, we observe that the UE positions are mostly in the correct zones (i.e., rooms and hallways), which demonstrates the capabilities of the bilateration and LoS bounding-box losses in D-MIMO scenarios with many APs.
Moreover, we observe that B2 is highly inaccurate, as we could not correct for the distortion in B1 through an affine transform no matter how many ground-truth labels we use to find such a transform.
This example demonstrates that channel charts may have non-affine distortions; hence, mapping channel charts to real-world coordinates through an affine transform is not a generally applicable method.

In \fref{tbl:indoor}, we observe---as expected---that B4 achieves the best performance in all metrics. 
P2 performs approximately the same as the semi-supervised baseline B3 in the latent space quality metrics, {by less than $0.009$ difference in TW and CT, and less than $0.043$ difference in KS and RD.
P2 has a slightly higher positioning error than B3, by approximately $0.2$\,m in mean and $0.8$\,m} in the 95th percentile.
P2 has a higher positioning error than B4 by approximately $0.5$\,m in mean and $1.3$\,m in the 95th percentile. 
This result demonstrates the efficacy of our proposed method P2 in an indoor area with many non-LoS paths.

\begin{figure*}[tp]
    \centering
    \subfigure[Ground-truth UE positions]
    {
    \includegraphics[width=\figsize\textwidth]{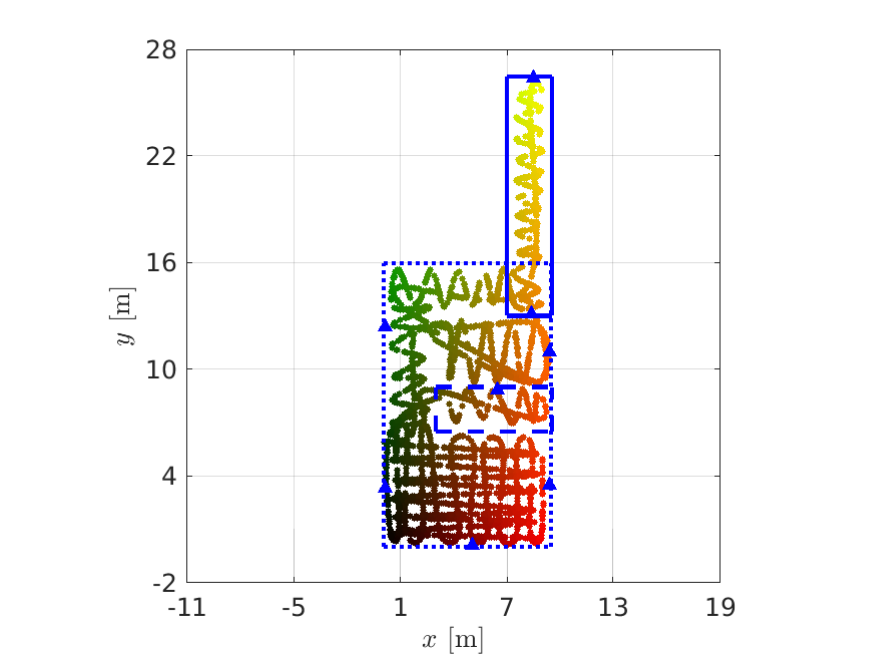}
    \label{fig:indoor_gt}
    }
    \subfigure[P1 (weakly-supervised)]  
	{
    \includegraphics[width=\figsize\textwidth]{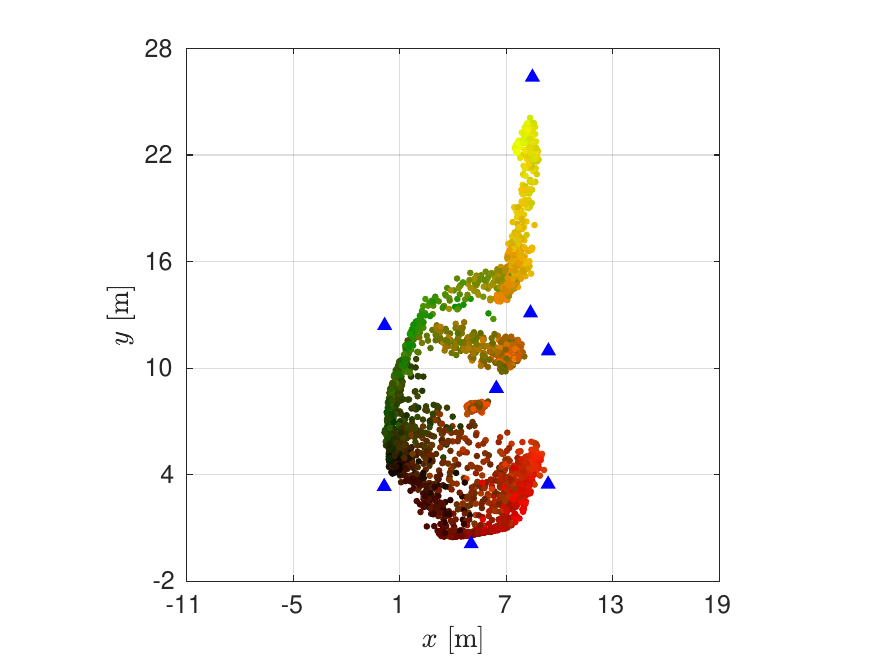}
    \label{fig:indoor_p1}
    }
    \subfigure[P2 (hybrid self- and weakly-supervised)]
	{
    \includegraphics[width=\figsize\textwidth]{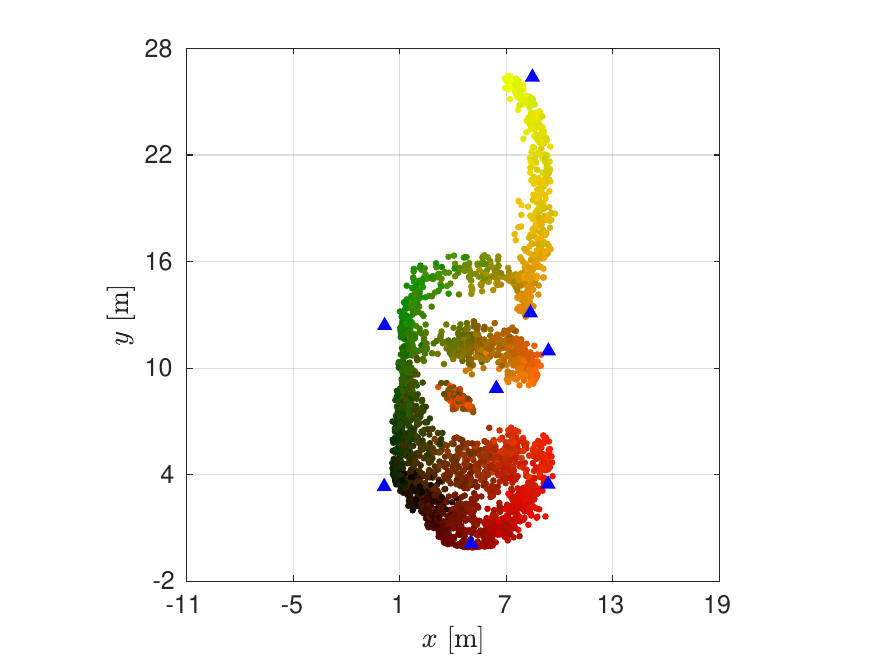}
    \label{fig:indoor_p2}
    }
     \subfigure[B1 (self-supervised)]
	{
    \includegraphics[width=\figsize\textwidth]{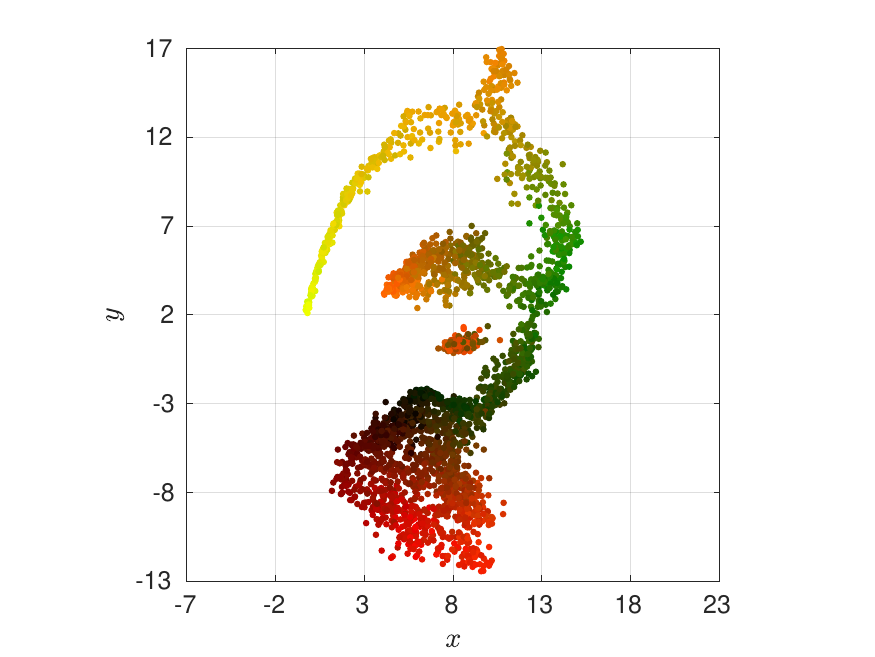}
    \label{fig:indoor_triplet}
    }
    \subfigure[B2 (affine transformed B1)]
	{
    \includegraphics[width=\figsize\textwidth]{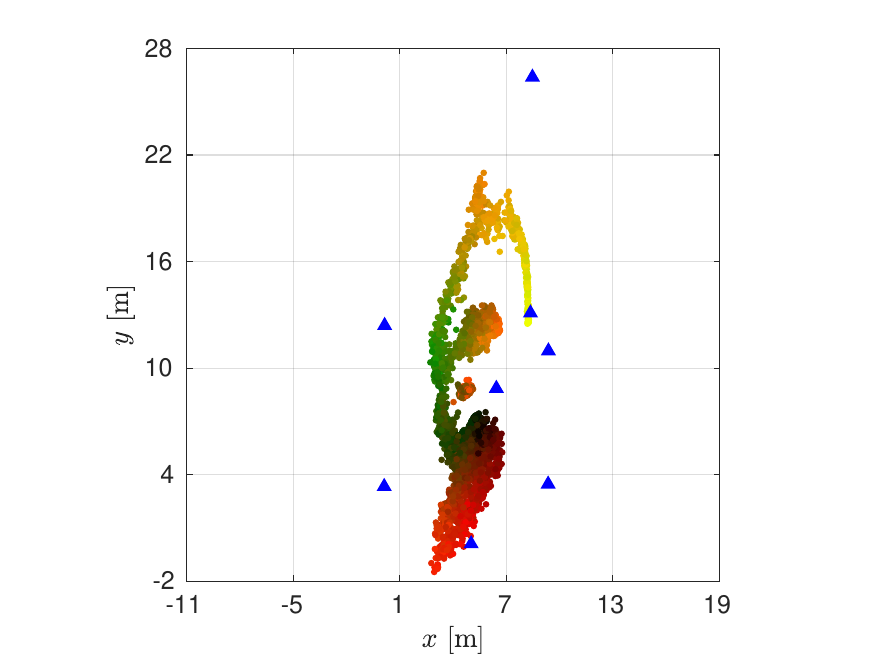}
    \label{fig:indoor_affine}
    }
    \subfigure[B3 (semi-supervised)]
	{
    \includegraphics[width=\figsize\textwidth]{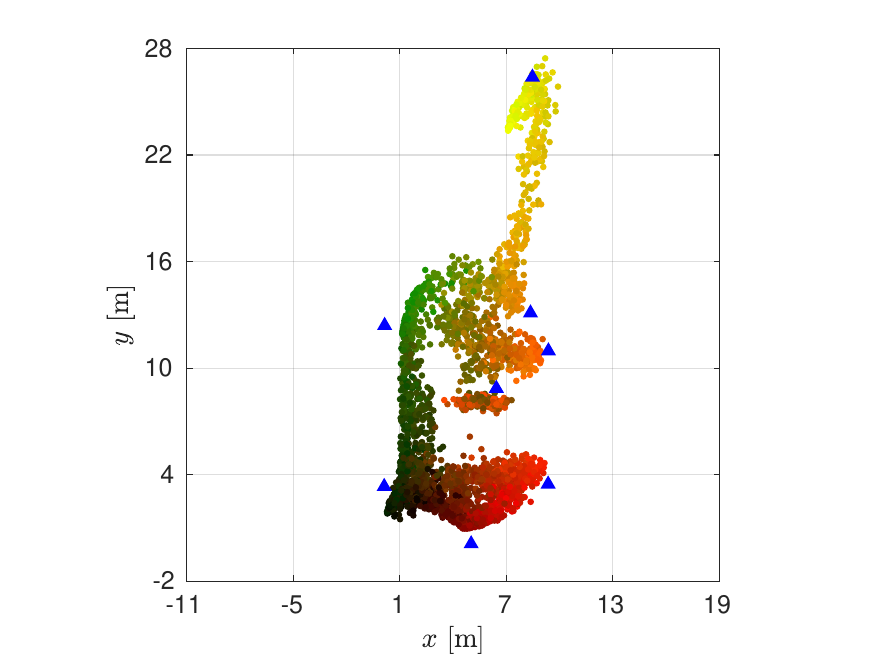}
    \label{fig:indoor_semisup}
    }
    \caption{Results for the simulated indoor scenario: (a) ground-truth UE positions (green-to-red gradient colored area), AP positions (blue triangles), and the LoS bounding box of three APs (blue solid and dashed lines); and (b-f) the channel charts or positioning estimates for the proposed (P) and baseline (B) methods. Since the output of baseline B2 is in arbitrary coordinates, the AP positions are not shown in (d). The proposed method P2 achieves comparable results in real-world coordinates as the semi-supervised baseline B3 but without requiring known UE positions during training.
    } 
    \label{fig:indoor_results}   
\end{figure*}

\begin{table*}
\centering 
\resizebox{0.92\textwidth}{!}{
\begin{minipage}[c]{0.99 \textwidth}
    \centering
    \caption{Channel charting and positioning performance comparison for the simulated indoor scenario}
    \label{tbl:indoor}		 
    \begin{tabular}{@{}lccccccc@{}}
        \toprule
        && \multicolumn{4}{c}{Latent space quality metrics} & \multicolumn{2}{c}{Positioning error [m]} \\
        \cmidrule(lr){3-6} \cmidrule(){7-8}  
        Method  & Figure & TW$\,\uparrow$ & CT$\,\uparrow$ & KS$\,\downarrow$ & RD$\,\downarrow$  &  Mean$\,\downarrow$ & 95th percentile$\,\downarrow$ \\
        \midrule
        P1 & {\ref{fig:indoor_p1}} & 0.950  & 0.952  & 0.195  & 0.791  & 1.61 $\pm$ 0.03 & 3.32 $\pm$ 0.12  \\  
        P2 & {\ref{fig:indoor_p2}} & 0.979  & 0.976  & 0.144 & 0.710  & 1.15 $\pm$ 0.04 & 2.84 $\pm$ 0.44 \\ 
        \midrule
        B1 & {\ref{fig:indoor_triplet}} & 0.983  & 0.968  & 0.347  & 0.857 & -- & -- \\  
        B2 & {\ref{fig:indoor_affine}} & 0.967  & 0.955  & 0.373  & 0.883 & 3.21 $\pm$ 0.54  & 6.66 $\pm$ 0.90  \\  
        B3 & {\ref{fig:indoor_semisup}} & 0.970  & 0.974  & 0.184  & 0.753 & 1.31 $\pm$ 0.18  & 3.53 $\pm$ 1.46  \\ 
        B4 &  --  & 0.988  & 0.989  & 0.082 & 0.572   & 0.63 $\pm$ 0.01   & 1.55 $\pm$ 0.02   \\
        \bottomrule
    \end{tabular}		 
\end{minipage}}
\end{table*}


\subsection{Measurement-Based Indoor Scenario}

\subsubsection{Description}
For our third scenario, we utilize the measured channel vectors from the DICHASUS datasets~\cite{dichasus2021,dataset-dichasus-cf0x}. The third column of~\fref{tbl:simulation_parameters} summarizes the measurement parameters.
These measurements are from a D-MIMO communication system in a factory environment, where the transmitter is a robot moving in various trajectories in an L-shaped area. 
The inner corner of this L-shaped area has a large metal container potentially blocking the LoS path to some APs, while every position in the measurement area has a LoS path to at least one AP;
see {\fref{fig:dichasus_gt}} for an illustration of the UE and AP positions along with the two LoS bounding boxes used for the top and bottom pairs of APs. 
We use the CSI from the first three trajectories of the dataset in~\cite{dataset-dichasus-cf0x} and take every fourth sample to reduce the total dataset size.
We apply a rotation of $12^\circ$ to all positions in order to simplify setting the rectangular LoS bounding boxes of APs.
Each CSI matrix $\bH^{(n)}$, $n\in\setN$ is of dimension $32\times 1024$; each CSI feature vector is of dimension $D'=32\times13=416$, where we use the $C=13$ taps  described in~\cite{stephan2023adp} and the feature extraction pipeline from \fref{sec:feature_extraction}. 

\subsubsection{Analysis of AP-Side Receive Power}
\label{sec:dichasus_power}

In this scenario, we know that not all UE positions have a LoS path to all APs; therefore, we first need to choose a power threshold to estimate the LoS APs to be used in bilateration and LoS bounding-box losses.
We choose this threshold based on the AP-side receive power $P^{(n,a)}$ over time $t_n$, as exemplified for one of the APs in {\fref{fig:dichasus_pow_time}}.
Here, we observe a steep drop in the power at around $15$\,dB, and set $\Pthr=15$. 
For brevity, we omit an ablation study that shows $\Pthr$ is robust to small deviations.
We now exemplify the power-distance relation exploited by our bilateration loss in \fref{eq:loss_bi} and the impact of $\Pthr$ for the same AP as~{\fref{fig:dichasus_pow_time}} in~{\fref{fig:dichasus_pow_distance}}.
Here, we observe the urgent need for choosing an appropriate power threshold $\Pthr$ as there are many UE positions that are close to the AP while the AP-side receive power is low;
we should not use this AP in bilateration loss for such UE positions. 
As desired, we observe that the receive power indeed falls below our chosen power threshold $\Pthr$ for these positions.
Moreover, the UE positions for which the AP-side receive power is higher than $\Pthr$  follow the same trend as \fref{fig:outdoor_pow_distance}, i.e., the power generally decreases with distance.
Therefore, we conclude that our choice of $\Pthr$ is suitable for classifying the LoS APs to be used in bilateration and LoS bounding-box losses.

For this CSI dataset, margins $\Mp$ of $0$, $3$, and $6$ result in false AP pair ratios of  $27.5$\%, $5$\%, and $1.8$\% for all UEs, 
while the average number of AP pairs in the set $\setP^{(n)}$ given by $\frac{1}{N}\sum_{n=1}^N |\setP^{(n)}|$ is $1.7$, $0.44$, and $0.10$, respectively.
Once again, we observe the trade-off in choosing~$\Mp$: Increasing $\Mp$ leads to fewer false AP pairs at the cost of fewer AP pairs in $\setP^{(n)}$ from \fref{eq:setP}.

\subsubsection{Parameter Settings}
For our proposed method P1, we set $ \Mt=\Mb=0.5$, $\Mp=3$, $\lambdat = 0$, and $\lambdabi = \lambdabox=1$.
P2 shares the same parameters as P1 except for the triplet loss, where we use $\lambdat = 1$, $\Tc=5$, and $\Mt=0.5$. B1, B2, and B3 use the same triplet loss parameters as P2.

For the baselines B2 and B3, we used the ground-truth positions of $10$ and $30$ CSI samples, respectively. With B2, we demonstrate the performance of learning an affine transform from a small number of labeled samples.
With B3, we demonstrate how many ground-truth labeled samples one would approximately need to achieve comparable performance to that of~P2.

\begin{figure}
\centering
\subfigure[AP-side receive power with respect to time]
{
\includegraphics[width=0.9\columnwidth]
{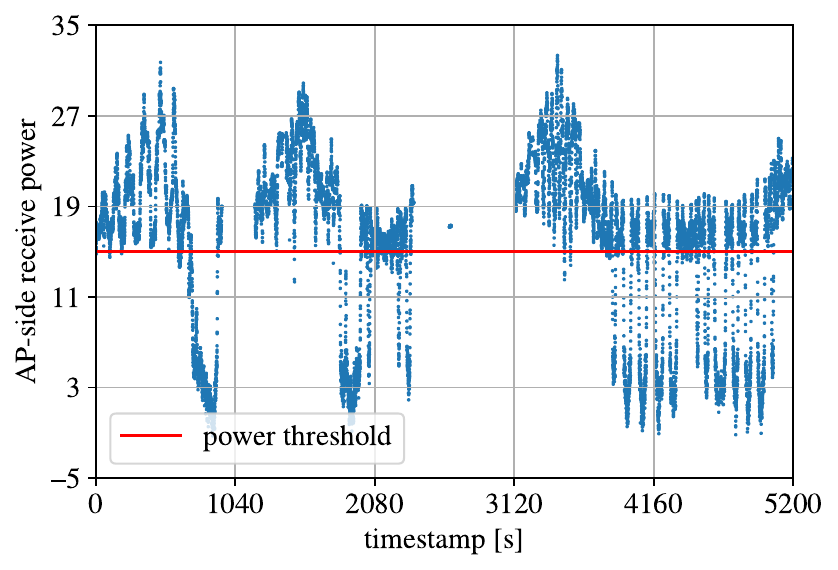}
\label{fig:dichasus_pow_time}
}
\subfigure[AP-side receive power versus AP-to-UE distance]
{
\includegraphics[width=0.9\columnwidth]{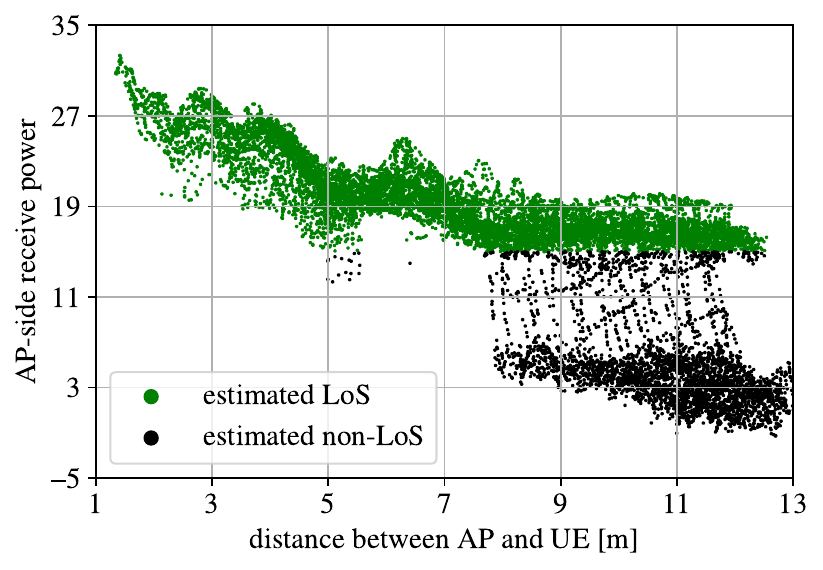}
\label{fig:dichasus_pow_distance}
}
\caption{Receive power (in decibels) at one AP for the measurement-based indoor scenario~\cite{dataset-dichasus-cf0x} with respect to (a) timestamps and (b) AP-to-UE distance in the xy-plane. In (a), the red line designates the power threshold chosen to identify LoS APs. In (b), the power values above the threshold (highlighted with green) indicate instances where the AP is estimated to be in LoS.} 
\label{fig:dichasus_pow}
\end{figure}

\subsubsection{Performance Evaluation}
\label{sec:dichasus_eval}

In \fref{fig:dichasus_results}, we observe 
that all methods that utilize triplet loss (i.e., P2, B1, B2, and B3) separate the red and green regions of the positions in an L-shaped chart resembling the ground-truth positions.   
However, for P1, the limitation of using bilateration and LoS bounding-box loss without the triplet loss is clear: 
With few APs and relatively large bounding boxes, the channel chart cannot reflect the UE positions well.

In \fref{tbl:dichasus}, we observe, as expected, that B4 achieves the best performance in all metrics. 
B2 is the second-best in all metrics as we again observe the advantage of starting from a ``good'' channel chart for the affine transform-based approach.
{Comparing the performance of P2 to B3, we observe that P2} has a slightly better performance in TW and CT {by less than $0.017$ and $0.11$}, and a slightly worse performance in KS and RD {by $0.010$ and $0.016$, respectively}.
P2 has a higher positioning error than B4 by approximately $1.2$\,m in mean and $2.6$\,m in the 95th percentile. 
P2 has a slightly higher positioning error than B2, by approximately $0.5$\,m in mean and $1.3$\,m in the 95th percentile; these differences are smaller for B3.
This result demonstrates the efficacy of our proposed method P2 for measured channels in an industrial environment, with few APs and many non-LoS paths.

\begin{figure*}[tp]
    \centering
    \subfigure[Ground-truth UE positions]
    {
    \includegraphics[width=\figsize\textwidth]{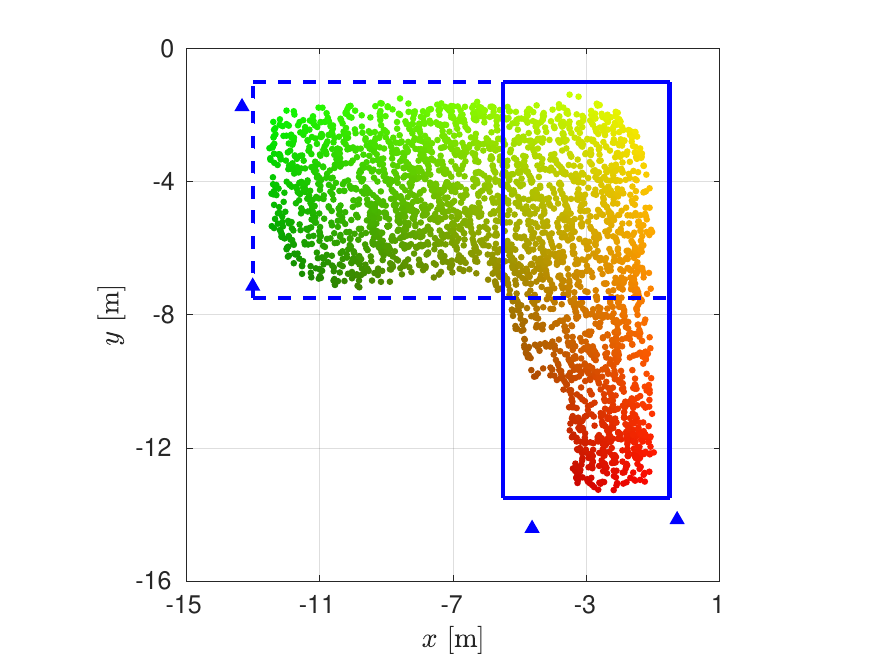}
    \label{fig:dichasus_gt}
    }
    \subfigure[P1 (weakly-supervised)]  
	{
    \includegraphics[width=\figsize\textwidth]{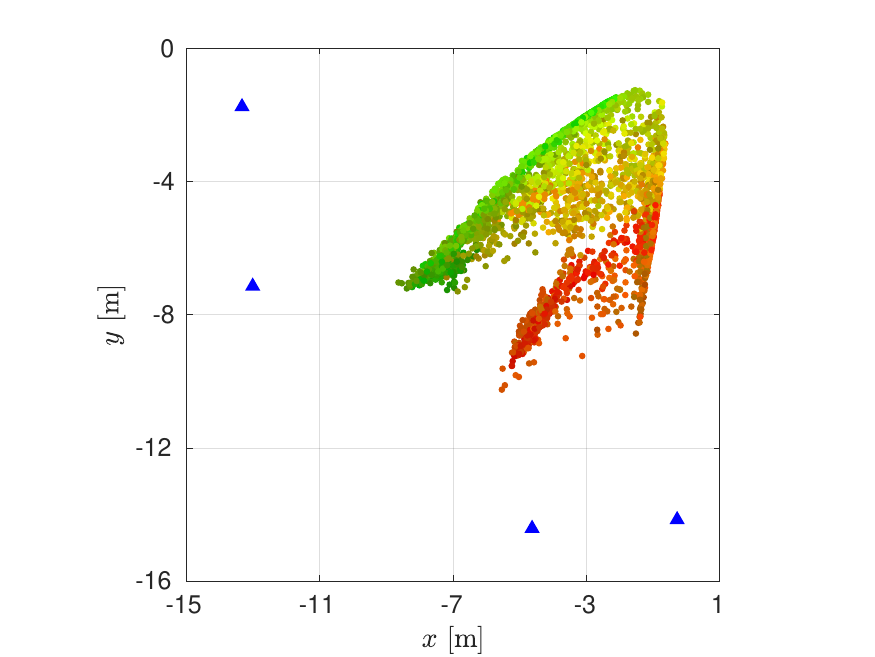}
    \label{fig:dichasus_p1}
    }
    \subfigure[P2 (hybrid self- and weakly-supervised)]
	{
    \includegraphics[width=\figsize\textwidth]{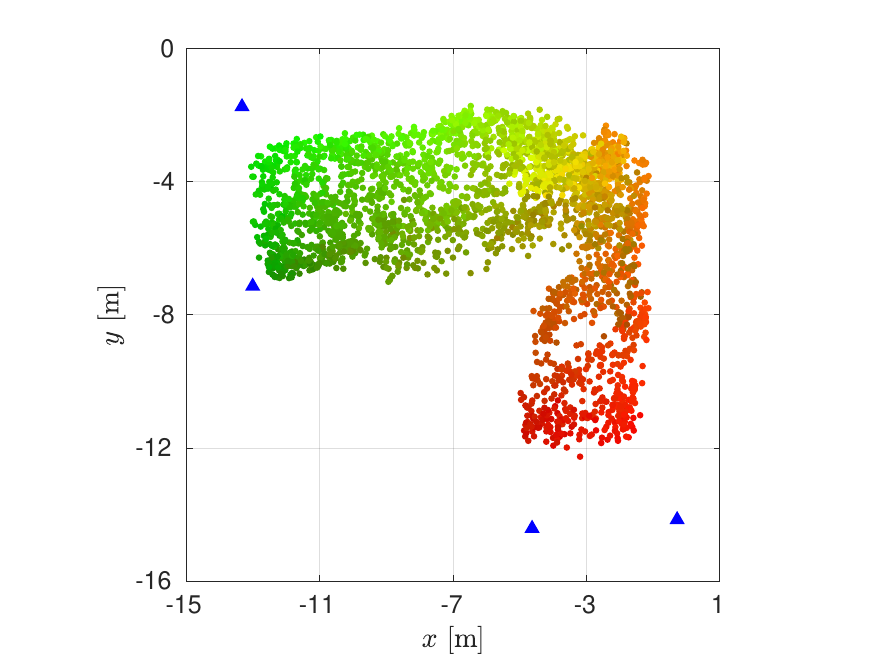}
    \label{fig:dichasus_p2}
    }
     \subfigure[B1 (self-supervised)]
	{
    \includegraphics[width=\figsize\textwidth]{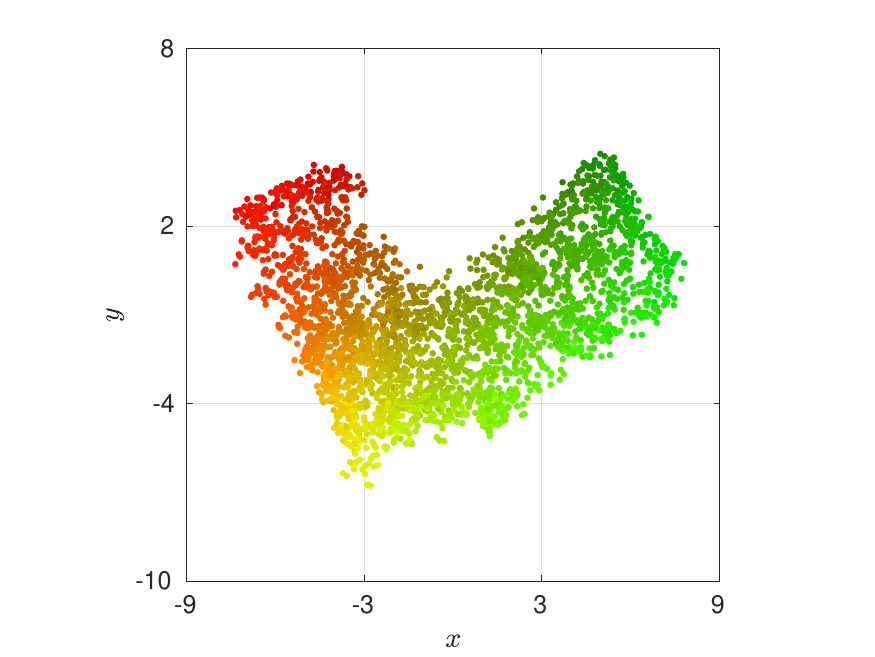}
    \label{fig:dichasus_triplet}
    }
    \subfigure[B2 (affine transformed B1)]
	{
    \includegraphics[width=\figsize\textwidth]{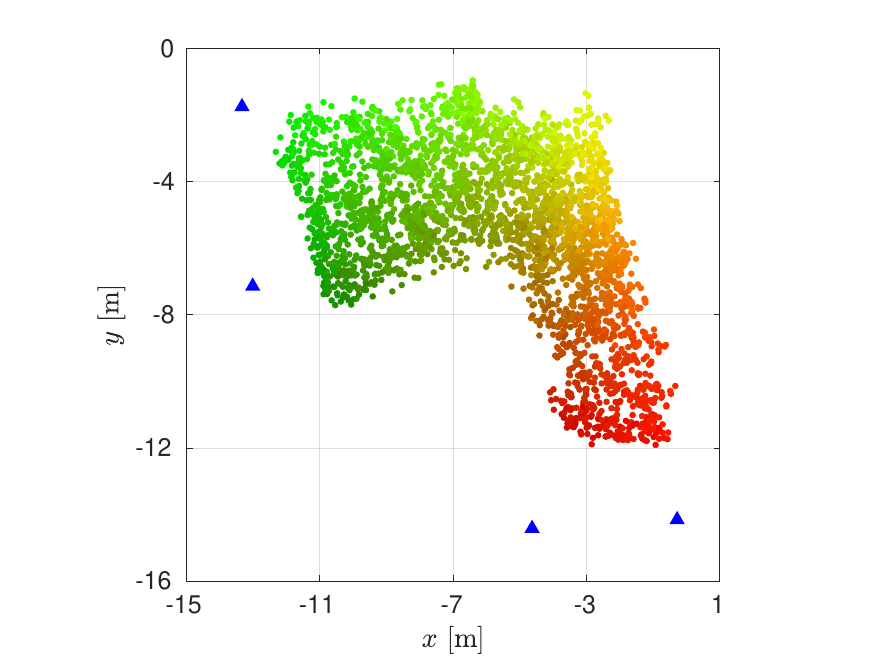}
    \label{fig:dichasus_affine}
    }
    \subfigure[B3 (semi-supervised)]
	{
    \includegraphics[width=\figsize\textwidth]{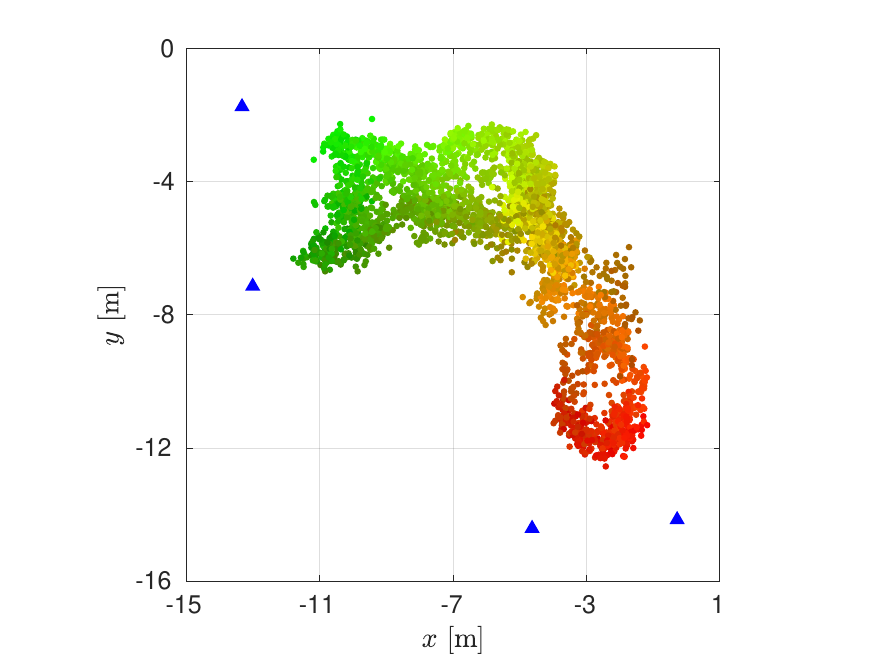}
    \label{fig:dichasus_semisup}
    }
    \caption{Results for the measured indoor scenario: (a) ground-truth UE positions (green-to-red gradient colored area), AP positions (blue triangles), and the two LoS bounding boxes corresponding to the top and bottom APs (blue solid and dashed lines); and (b-f) the channel charts or positioning estimates for the proposed (P) and baseline (B) methods. Since the output of baseline B2 is in arbitrary coordinates, the AP positions are not shown in (d). The proposed method P2 achieves comparable results in real-world coordinates as the semi-supervised baseline B3 but without requiring known UE positions during training.} 
    \label{fig:dichasus_results}   
\end{figure*}

\begin{table*}
\centering
\resizebox{0.92\textwidth}{!}{
\begin{minipage}[c]{0.99 \textwidth}
    \centering
    \caption{Channel charting and positioning performance comparison for the measured indoor scenario}
    \label{tbl:dichasus}		 
    \begin{tabular}{@{}lccccccc@{}}
        \toprule
        && \multicolumn{4}{c}{Latent space quality metrics} & \multicolumn{2}{c}{Positioning error [m]} \\
        \cmidrule(lr){3-6} \cmidrule(){7-8}  
        Method  & Figure & TW$\,\uparrow$ & CT$\,\uparrow$ & KS$\,\downarrow$ & RD$\,\downarrow$  &  Mean$\,\downarrow$ & 95th percentile$\,\downarrow$ \\
        \midrule
        P1 & {\ref{fig:dichasus_p1}} & 0.842  & 0.852  & 0.441  & 0.961  & 2.83 $\pm$ 0.24  & 6.28 $\pm$ 0.61   \\  
        P2 & {\ref{fig:dichasus_p2}} & 0.972  & 0.980  & 0.192  & 0.799 & 1.33 $\pm$ 0.03 & 3.05 $\pm$ 0.17 \\ 
        \midrule 
        B1 & {\ref{fig:dichasus_triplet}} & 0.989  & 0.990  & 0.179  & 0.769 & -- & -- \\  
        B2 & {\ref{fig:dichasus_affine}} & 0.990  & 0.991  & 0.125  & 0.678 & 0.83 $\pm$ 0.37  & 1.78 $\pm$ 0.76  \\  
        B3 &  {\ref{fig:dichasus_semisup}}  & 0.955  & 0.969  & 0.182  & 0.783 & 1.02 $\pm$ 0.06  & 2.47 $\pm$  0.30 \\ 
        B4 &  --  & 0.998  & 0.998  & 0.036  & 0.401 & 0.18 $\pm$ 0.01  & 0.42  $\pm$ 0.02   \\
        \bottomrule
    \end{tabular}		 
\end{minipage}}
\end{table*}

\subsection{{Comparing the Results of Our Three Scenarios}}
\label{sec:compare3scenarios}

{
By evaluating the proposed methods and baselines over three scenarios, we have observed that our weakly-supervised method P2 is able to achieve similar performance as the semisupervised baseline method B3. By comparing the results between our three scenarios, we observe the following:}
\begin{itemize}
    \item {The number of ground-truth UE positions required in the baseline method B3 to achieve similar performance to our proposed P2 depends on the scenario and the size of the area of interest. 
    For example, our simulated outdoor scenario, in which the UEs were placed in a much larger area than in the two indoor scenarios, required $400$ samples, whereas the indoor scenarios required at most~$50$~samples.}

    \item {The performance gap between the proposed method P2 and the fully-supervised baseline B4 is  larger in our outdoor scenario than in the two indoor scenarios.
    {We attribute the large performance gap between P2 and B4 in the outdoor scenario to the large size of the LoS bounding box, which is approximately equal to the entire area of UE positions
    since all APs have LoS paths to all UE positions.}
    In contrast, {the APs in our indoor scenarios do not have LoS paths to many UE positions and have small LoS bounding boxes}  that divide the total area of UE positions. Hence,
    placing a UE in one of these small boxes already significantly reduces the positioning error and reduces the gap between P2 and B4.
    }
\end{itemize}
{We hope that these observations will guide future work including even more diverse scenarios.}



\section{Conclusions and Future Work}
\label{sec:conclusions}

We have proposed the bilateration loss and LoS bounding-box loss which enable weakly-supervised channel charting in real-world coordinates, without requiring geometric models, accurate AP synchronization, or ground-truth UE positions.
The bilateration loss utilizes the known AP positions in a D-MIMO scenario by comparing the received power at pairs of APs that are estimated to have a LoS path to the UE and places the UE closer to the AP which receives higher power.
The LoS bounding-box loss places the UE in the bounding box of each AP that is estimated to have a LoS path.
We have demonstrated that a multi-loss combining the bilateration and LoS bounding-box losses with the triplet loss from~\cite{ferrand2021} is sufficient to generate high-quality channel charts in real-world coordinates
using three different datasets: two datasets based on a commercial ray tracer (one indoors and one outdoors) and one based on indoor channel measurements.
We have shown through a comparison with several baselines that our proposed approach can achieve the performance of using a subset of ground-truth labeled CSI samples, which has the advantage that no ground-truth positions need to be acquired. 
In addition, the size of the subset of known CSI samples for semi-supervised baseline methods depends heavily on the scenario and the size of the area of interest.
Moreover, we have demonstrated that a two-stage approach that maps a learned channel chart to real-world coordinates using an affine transform may completely fail depending on the channel chart quality.

There are many avenues for future work. 
In order to improve positioning accuracy, the bounding boxes could be extended to more general shapes, the LoS/non-LoS channel classification could be performed using more advanced (and automated) methods than power-thresholding, and
improved loss functions that compare AP angle-of-arrival information besides receive power could be utilized as well.
In addition, investigating the efficacy of other dissimilarity metrics further to improve the quality of the learned channel charts is part of ongoing work.


\bibliographystyle{IEEEtran}

\balance

\bibliography{bib/IEEEabrv,bib/confs-jrnls,bib/publishers,bib/studer,bib/vipbib, bib/cc_bib, bib/emres_bib} 

\balance

\end{document}